\documentclass[lettersize,journal]{IEEEtran}
\usepackage{comment}
\usepackage{cite}
\usepackage{amsmath,amssymb,amsfonts}
\usepackage{fixltx2e}
\usepackage{graphicx}
\usepackage{textcomp}
\usepackage{xcolor}
\usepackage{comment}
\usepackage{multirow}
\usepackage{subcaption}
\usepackage{balance}
\usepackage{algpseudocode}
\usepackage[utf8]{inputenc}
\usepackage{amsmath}
\usepackage{amsfonts}
\usepackage{amssymb}
\usepackage[ruled,vlined,linesnumbered]{algorithm2e}
\usepackage{hyperref}
\usepackage{algpseudocode}
\usepackage[linesnumbered,ruled,vlined]{algorithm2e}
\usepackage{nomencl}
\usepackage{fancyhdr}
\usepackage{lipsum} 
\fancypagestyle{accepted}{
    \fancyhf{}
    
    \fancyfoot[L]{\footnotesize \copyright 2024 IEEE. Personal use of this material is permitted. Permission from IEEE must be obtained for all other uses, in any current or future media, including reprinting/republishing this material for advertising or promotional purposes, creating new collective works, for resale or redistribution to servers or lists, or reuse of any copyrighted component of this work in other works. 
    DOI: \href{https://doi.org/10.1109/TPDS.2024.3430853}{10.1109/TPDS.2024.3430853}
    }
}

\begin{document}

\title{STT-RAM-based Hierarchical In-Memory Computing
}

\author{Dhruv Gajaria,~\IEEEmembership{Member,~IEEE}, Kevin Antony Gomez, and Tosiron Adegbija,~\IEEEmembership{Senior Member,~IEEE}
\thanks{Gajaria and Adegbija are with the Department of Electrical and Computer Engineering, The University of Arizona, USA, email: \{dhruvgajaria\}@arizona.edu, \{tosiron\}@arizona.edu. Gomez is with the Department of Computer Science, University of Massachusetts, Amherst, USA, email: \{kantonygomez\}@umass.edu This work was done when Kevin Gomez was at the University of Arizona.}
\thanks{This work is partly supported by the National Science Foundation under grant CNS-1844952.}
}

\maketitle
\thispagestyle{accepted}
\begin{abstract}
In-memory computing promises to overcome the von Neumann bottleneck in computer systems by performing computations directly within the memory. Previous research has suggested using \textit{Spin-Transfer Torque RAM (STT-RAM)} for in-memory computing due to its non-volatility, low leakage power, high density, endurance, and commercial viability. This paper explores \textit{hierarchical in-memory computing}, where different levels of the memory hierarchy are augmented with processing elements to optimize workload execution. The paper investigates processing in memory (PiM) using non-volatile STT-RAM and processing in cache (PiC) using volatile STT-RAM with relaxed retention, which helps mitigate STT-RAM's write latency and energy overheads.
We analyze tradeoffs and overheads associated with data movement for PiC versus write overheads for PiM using STT-RAMs for various workloads. We examine workload characteristics, such as computational intensity and CPU-dependent workloads with limited instruction-level parallelism, and their impact on PiC/PiM tradeoffs. Using these workloads, we evaluate computing in STT-RAM versus SRAM at different cache hierarchy levels and explore the potential of heterogeneous STT-RAM cache architectures with various retention times for PiC and CPU-based computing. Our experiments reveal significant advantages of STT-RAM-based PiC over PiM for specific workloads. Finally, we describe open research problems in hierarchical in-memory computing architectures to further enhance this paradigm.
\end{abstract}


\section{Introduction}
The exponential growth of data in recent years has created a pressing need for efficient data processing in resource-constrained systems, such as mobile devices, Internet of Things (IoT) devices, and embedded systems. In-memory computing, or processing in memory (PiM), has the potential to meet this need, offering low-latency, high-throughput data processing capabilities that can operate within the constraints of these devices' limited resources. In-memory computing involves memory devices designed with processing units across memory arrays\cite{sebastian2020memory}, thereby preventing the need for costly data transfers between the processor and memory.

In-memory computing has gained traction and has the potential to significantly impact multiple domains, including scientific computing, healthcare, machine learning, autonomous driving, and more, thanks to its massive parallelism and reduced data movement. Previous studies have focused primarily on using in-memory processing units as accelerators for kernel execution. However, in many real-world applications, both the processor and in-memory computing are required to complete the necessary computations \cite{boroumand2018google} effectively. This may result in additional data movement overhead across the memory hierarchy \cite{compute_caches}. Address translation poses another challenge for in-memory computing, as the address spaces of main memories may exceed the translation lookaside buffer (TLB) capacity, leading to frequent and costly page walks \cite{address_translation_issue}. As a result, the applicability of in-memory computing may be limited in certain scenarios.

Previous studies have suggested using \textit{compute caches} \cite{compute_caches} or implementing \textit{processing in cache (PiC)} architectures. Specifically, they explored bit-line computing, which involves activating multiple word-lines in an SRAM cache to sense the resulting voltage and perform computations on the data. Bit-line computing allows the compute units to be placed in subarrays. This results in faster access latencies and more parallel units with simpler cache extensions than near-memory or DRAM-based in-memory computing \cite{das2017blurring}. However, implementing a PiC-based bit-line with SRAM may not always be practical, given the high power and area requirements of SRAM caches, along with the added overhead of integrating processing units with the cache. This is particularly true for resource-constrained systems. Furthermore, reducing the word-line voltage to prevent data corruption in SRAM during bit-line computing can increase the cache delay \cite{compute_caches, jeloka201628}.  

Spin-Transfer Torque RAM (STT-RAM) is an emerging non-volatile memory (NVM) technology that offers a promising alternative to SRAM for cache implementation. STT-RAM boasts a much smaller area requirement (40\% - 80\% less area) and negligible leakage power \cite{sun2011multi} compared to SRAM. STT-RAM has also demonstrated commercial viability and superior endurance compared to other NVM technologies, such as ReRAM or SOT-RAM, positioning it as a leading candidate to replace SRAM. Additionally, STT-RAM inherently has a higher write current than read current, preventing data corruption during bit-line computing \cite{jain2017computing}. Despite its promising features, STT-RAM suffers from high write latency and energy consumption due to its non-volatile nature. Researchers have proposed mitigating this challenge by relaxing the retention time---the duration for which data is retained in the cache in the absence of power---to satisfy only the retention time requirements of the cache blocks relevant to the executing workloads \cite{kuan2019energy}. To this end, the retention time can potentially be reduced to less than 1$s$ \cite{smullen2011relaxing, sun2011multi}, leading to substantial latency and energy savings.

In this paper, we propose and explore the idea of \textit{hierarchical in-memory computing} from the STT-RAM perspective. Hierarchical in-memory computing (HiMC) is an in-memory computing paradigm in which multiple levels of the memory hierarchy are augmented to enable a combination of PiC and PiM. As such, computations occur in the layer where it makes the most sense concerning the design objectives (performance or energy consumption), and data movement is minimized. We propose and study an HiMC system in which PiC is implemented using relaxed retention STT-RAM caches, and PiM is implemented using non-volatile STT-RAM. 

Unlike prior STT-RAM-based in-memory computing research, which has only considered non-volatile STT-RAM PiM implementations \cite{jain2017computing, parveen2018hielm}, our work for the first time considers the tradeoffs of computing across the memory hierarchy. Relaxed retention STT-RAMs introduce new considerations, like the impact of cache block lifetimes, that must be considered to maximize the energy benefits of STT-RAM caches. Our study explores different types of workloads to reveal the implications of various workload characteristics on computation efficiency. These characteristics include CPU dependence (due to sequential executions and unsupported PiC/PiM computations), temporal locality/data reuse, and write intensity, all of which have important implications for the effectiveness of PiC vs. PiM. We show the kinds of workloads best suited for PiC/PiM, propose solutions to potential PiC design issues, and analyze the latency and energy benefits of the proposed solutions compared to the state-of-the-art. To gain the benefits of both PiC and traditional CPU-based computing, we also explore for the first time, heterogeneous retention times caches featuring different retention times for PiC and CPU-based computing. 

In summary, we make the following key contributions:
\begin{itemize}
    \item For the first time, we study STT-RAM-based PiC with relaxed retention at different cache hierarchy levels (specifically, L1 and L2).  
    
    \item We investigate various types of workloads to assess the tradeoffs associated with computing at different levels of the memory hierarchy, including the cache hierarchy and main memory, using STT-RAM. Our analysis takes into account different workload characteristics that have significant implications for the effectiveness of PiC/PiM.

    \item We compare the energy and latency advantages of PiC using STT-RAM versus SRAM. The experimental results indicate that STT-RAM achieves substantial area savings (up to 79.86\% compared to SRAM) and energy savings (averaging 4.18x compared to CPU and up to 6\% [up to 54.5\%] compared to SRAM) while maintaining similar latency to SRAM-based PiC (3x improvement over CPU).
    \item To enhance the execution of CPU-dependent workloads, which face the most significant limitations with PiC/PiM, we propose a simple optimization technique called "operation chaining" that enhances the concurrency of execution between CPU and PiC/PiM units. This approach yields an average latency improvement of 10.19\% and energy savings of 7.83\% compared to the current state-of-the-art.
    \item To maximize the benefits of STT-RAM for both PiC and traditional CPU-based computing, we explore a novel heterogeneous cache design for PiC that uses lower retention times for PiC computing and higher retention times for CPU-based computing.
\end{itemize}
In the rest of the paper, Section \ref{sec: background} presents the background information on the STT-RAM cell and then discusses the relaxed retention STT-RAM caches and related work on STT-RAM-based PiC/PiM computing. Section \ref{sec:PiCdesign} presents the architecture for STT-RAM-based PiC caches, and their operations and scalability. The section also discusses finding the optimal retention time for PiC computing and introduces \textit{operation chaining} to optimize CPU-dependent workloads. Section \ref{sec:hetdesign} introduces heterogeneous STT-RAM caches for PiC and CPU-based computing. Finally, Section \ref{sec:exp} discusses our experimental setup, and Section \ref{sec:results} presents the results of our experiments and the overhead of the explored designs.

\section{Background and Related Work}\label{sec: background}
In this section, we first present an overview of the STT-RAM cell. We then present the tradeoffs of STT-RAM caches and discuss the previous work on STT-RAM caches. Following this, we discuss the prior work on PiC implementations using SRAMs, and an overview of the state-of-the-art PiC/PiM using STT-RAMs.

\subsection{Overview of the STT-RAM cell}

\begin{figure}[t]
	
		\centering
		\includegraphics[width=0.75\linewidth]{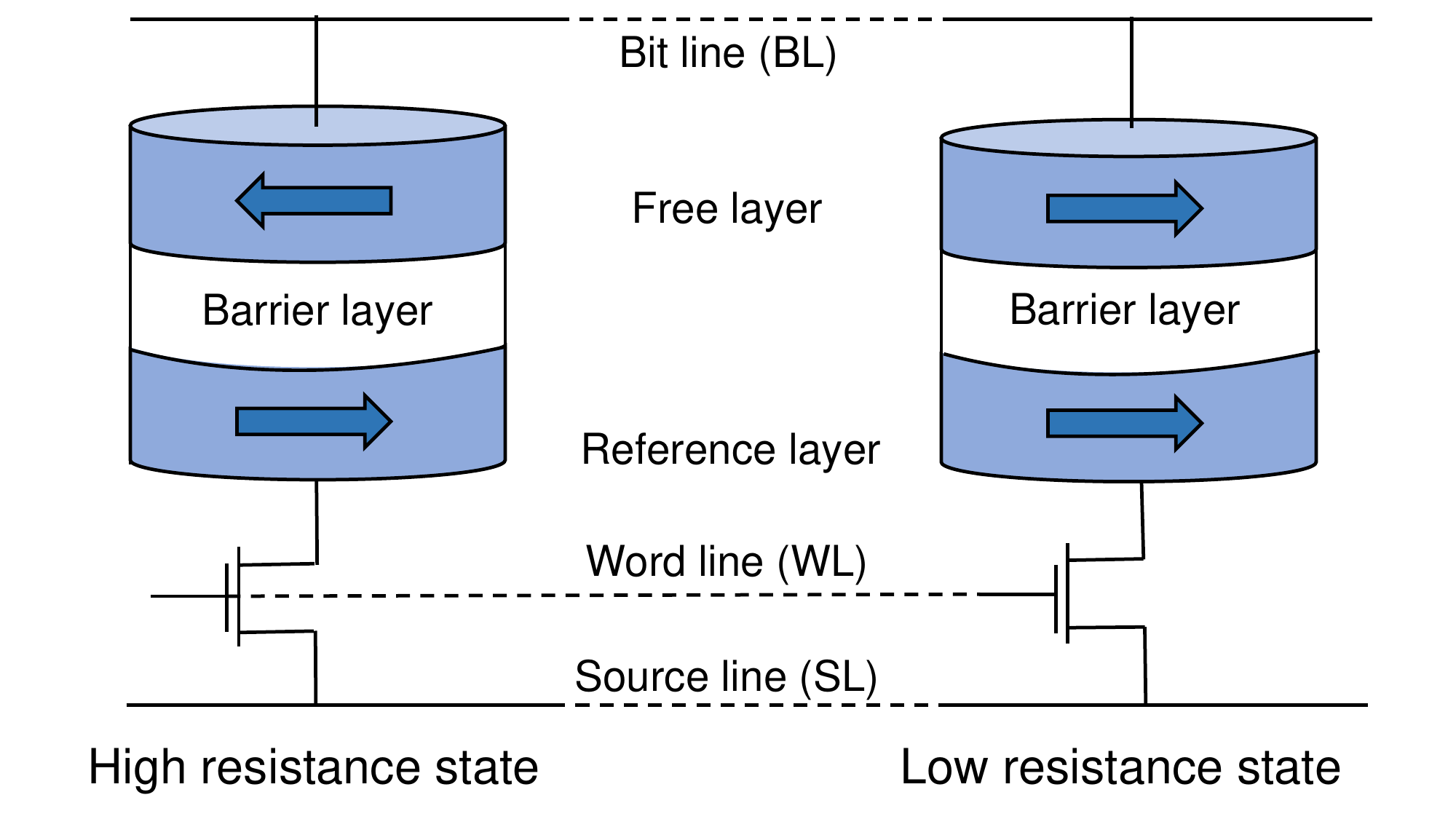}
		\caption{STT-RAM cell structure. The high resistance state is anti-parallel, while the low resistance state is parallel.}
		\label{fig:sttram_cell}
\vspace{-12pt}
\end{figure}

The cell structure of STT-RAM comprises a transistor and a magnetic tunnel junction (MTJ). The MTJ includes an oxide layer positioned between two ferromagnetic layers: the free layer and the hard (or fixed) layer. While passing a current through the free layer can alter its magnetization, the magnetization of the hard layer remains unchanged. The orientation of magnetization between these ferromagnetic layers determines the stored bit in the STT-RAM cell: '0' (in the parallel state or low resistance state $R_{P}$) or '1' (in the anti-parallel state or high resistance state $R_{AP}$), as illustrated in Figure \ref{fig:sttram_cell}. Further insights into the characteristics of $R_{P}$ and $R_{AP}$ and the functioning of STT-RAM cells can be found in previous research \cite{chun2012scaling}.

STT-RAM exhibits several attributes, such as high area density, extremely low leakage power, superior endurance (in comparison to other non-volatile memories \cite{wang2013low}), and excellent scalability \cite{sun2011multi}. STT-RAM occupies around 40-80\% of the area required by SRAM, enabling the provision of larger caches and accommodating smaller area constraints for PiC computing within the same die footprint. Nevertheless, deploying STT-RAM in caches presents challenges due to its high write latency and energy requirements. These high overheads make adaptation of STT-RAM difficult in caches. To mitigate these overheads, Smullen et al. \cite{smullen2011relaxing} proposed the relaxation of retention time in STT-RAM caches. The researchers discovered that the retention time of STT-RAM is exponentially proportional to the energy associated with magnetization stability, which can be expressed as:
\begin{equation*}
\Delta \propto \frac{V \cdot H_{k} \cdot M_{s}}{T}
\end{equation*}
Here, $V$ represents the activation volume for the writing current of STT-RAM, $H_{k}$ denotes the in-plane anisotropy field, $M_{s}$ indicates the saturation magnetization and $T$ represents the absolute temperature in Kelvin \cite{smullen2011relaxing}. The following section will explore how prior studies leveraged the relaxation of retention time in STT-RAM caches.

\subsection{Relaxed-retention STT-RAM caches}
Prior work by Smullen et al. \cite{smullen2011relaxing} found that the retention time can be relaxed by reducing the area of the cell to mitigate the tasking write operations for STT-RAM caches. However, this design is less efficient in high switching regions (latency$<$3ns) \cite{sun2011multi}. Thus, prior work by Sun et al. \cite{sun2011multi} found that the retention time can be relaxed by changing the thickness of the free layer, magnetization saturation (M\textsubscript{s}), and the effective anisotropy field (H\textsubscript{k}). Moreover, since the average cache block lifetime is less than one second, the authors explored relaxed retention time for different levels of cache hierarchy. The authors also proposed a multi-level retention time STT-RAM cache hierarchy that achieves average energy savings of 73.8\% over SRAM cache designs while maintaining the same instructions per cycle (IPC) performance. Thus, a relaxed retention STT-RAMs can achieve significant energy benefits compared to SRAMs throughout the cache hierarchy. In this study, we use a technique similar to \cite{sun2011multi} to model the relaxed retention STT-RAM caches.

Kuan et al. \cite{kuan2019energy} discovered that energy-efficient optimization of relaxed retention STT-RAM caches can be achieved by closely aligning the configuration of retention time with the runtime execution demands of workloads. The researchers introduced a logically adaptable retention STT-RAM (LARS) L1 cache, which incorporates multiple units of retention time. They utilized a sampling-based algorithm to determine the most suitable retention times for different applications dynamically.
Similarly, Gajaria et al. \cite{gajaria2019scart} found that specializing retention times of STT-RAM caches to the application needs improved performance and energy savings by 20.34\% and 29.12\%, respectively.
In our work, we explore how relaxed retention STT-RAM can improve the efficiency of PiC operations. We also explore heterogeneous STT-RAM cache design schemes for PiC and CPU-based computing. 

\subsection{STT-RAM-based processing in cache/memory (PiC/PiM)}

Bit-line computing is more suitable for non-volatile memories and SRAM than DRAM-based memory due to frequent refreshes required by DRAMs \cite{das2017blurring, ankit2020circuits}. To ensure reliable bit-line operations the source operands need to be fully charged. These prior works ensured that the source operands are refreshed before any operations \cite{seshadri,gao2019computedram}.  Thus, previous works by Seshadri et al. \cite{seshadri}, and Gao et al. \cite{gao2019computedram} proposed copying source operands to reserved rows to perform bit-line computing, ensuring that the source operands are always refreshed. 
However, unlike DRAMs, non-volatile memories do not have to move the data to reserved rows.

Many previous studies have investigated processing in memory (PiM) utilizing non-volatile memories (NVM) \cite{jain2017computing, li2016pinatubo, fan2017memory}. For instance, Fan et al. \cite{fan2017memory} proposed an in-memory AES accelerator employing spintronic devices, demonstrating a 58.6\% reduction in energy consumption compared to CMOS circuits. Li et al. \cite{li2016pinatubo} explored bitwise PiM utilizing NVMs like phase-change memory (PCM) and resistive RAM (ReRAM) for data-intensive applications, achieving a 500x speedup for bitwise operations and an overall speedup of 1.12x for data-intensive graph applications.

Parveen et al. \cite{parveen2018hielm} proposed an STT-RAM-based in-memory computing architecture that performed Boolean logic operations on any two cells of the same memory array. They evaluated their work in-memory Boolean vector logic and found optimization by 8x and 5x for energy and latency over DRAM-based in-memory computing. 
Jain et al. proposed using bit-line computing with STT-RAM memories featuring a compute unit of AND, OR, NOR, NAND, NOT, XOR, and ADD operations. They also evaluated a reliable PiM implementation under process variations with STT-RAMs. Their architecture achieves system-level performance and energy improvements by 3.93x and 3.83x, respectively.
Their work, which we leverage for our PiM implementation, features an STT-RAM PiM design that performs addition and logical operations.

Although PiC (Processing in Cache) has not received as much attention as PiM (Processing in Memory), previous research \cite{compute_caches} has demonstrated that PiC can address data transfer overheads that may arise in certain workload types for PiM. Aga et al. \cite{compute_caches} proposed integrating computing units within SRAM caches to mitigate data transfer overhead, leading to performance improvements and energy reduction by factors of 1.9x and 2.4x, respectively, compared to CPU-only computing. Eckert et al. \cite{eckert2018neural} introduced a Neural Cache that transforms SRAM caches into parallel compute units specifically designed for deep neural network inference. Nag et al. \cite{nag2019gencache} proposed GenCache, a method utilizing SRAM-based in-cache computing to accelerate the genetic sequence alignment task. Our work \cite{gajaria2022study} is the first to explore PiC using STT-RAM and analyze the tradeoffs associated with STT-RAM-based computing at various memory hierarchy levels.

\section{Overview of hierarchical in-memory computing}

\begin{figure}[t]
	
		\centering
		\includegraphics[width=0.65\linewidth]{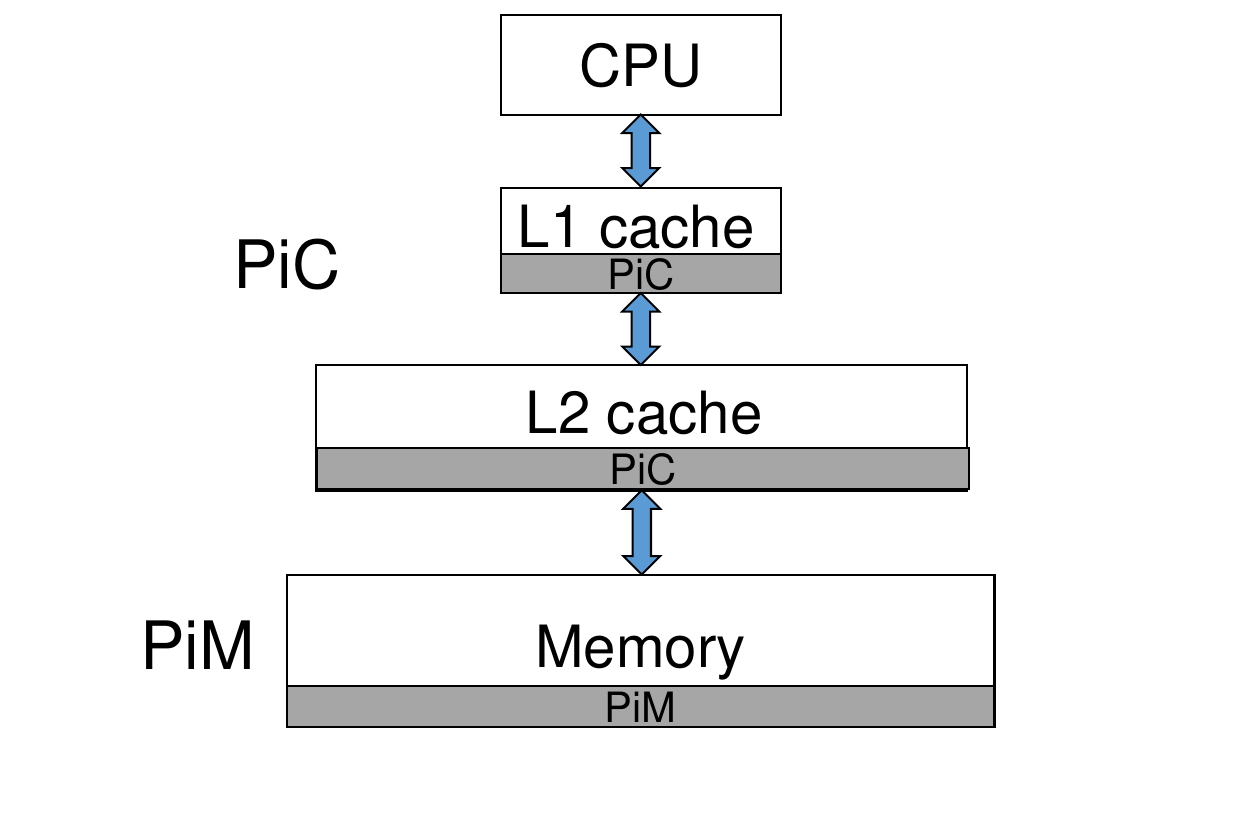}
		\caption{The system model featuring processing in cache (PiC) implemented in the L1 and L2 caches and processing in memory (PiM) implemented in the main memory.}
		\label{fig:system}
\vspace{-12pt}
\end{figure}

Before discussing the PiC architecture, we first present an overview of hierarchical computing. Figure \ref{fig:system} illustrates the system model considered in our work. The general idea of hierarchical in-memory computing is to perform computations where it makes the most sense---as close to the data as possible. As such, the relaxed retention STT-RAM L1 and L2 caches are augmented with computing circuits to enable PiC, while the non-volatile STT-RAM main memory is augmented to enable PiM. This paper focuses on the tradeoffs of computing at different levels of the memory hierarchy, considering the data movement overheads.

The L1 cache is closest to the CPU and has the fastest access and compute latency. But since L1 is smaller, it cannot feature a large number of concurrent compute units without inducing significant penalties on the CPU's regular cache accesses. On the other hand, PiC-based L2 caches will have longer data access times but can have a larger number of compute units than L1 caches. Similarly, PiM will have the highest access latency but can have the highest number of compute units due to its larger area. 
Furthermore, PiM comprises non-volatile STT-RAMs with far greater write overheads than the relaxed-retention STT-RAMs in the PiC. We will discuss the architecture of the compute units in detail in Section \ref{sec:arch}.

Considering the data movement overheads, PiM and PiC will have different tradeoffs depending on the location of operands and the availability of compute units. PiM will generally have the lowest data overheads for computations on data resident in the memory, allowing data to be processed without having to communicate with the processor. Alternatively, PiC is generally faster and will have the least data movement overheads for the data that are either present in the cache or originate from the CPUs (e.g., through stores). In general, the choice between PiC and PiM will depend on the specific application requirements and the underlying architecture of the compute units. In this paper, we explore these tradeoffs with respect to CPU-dependent workloads to evaluate CPU-based data movement, workloads with high reuse data to evaluate PiC vs PiM write overheads, and CPU-independent workloads.

\section{Relaxed Retention Processing in Cache (PiC)}\label{sec:PiCdesign}

\begin{figure}[t]
		\centering
		\includegraphics[width=0.99\linewidth]{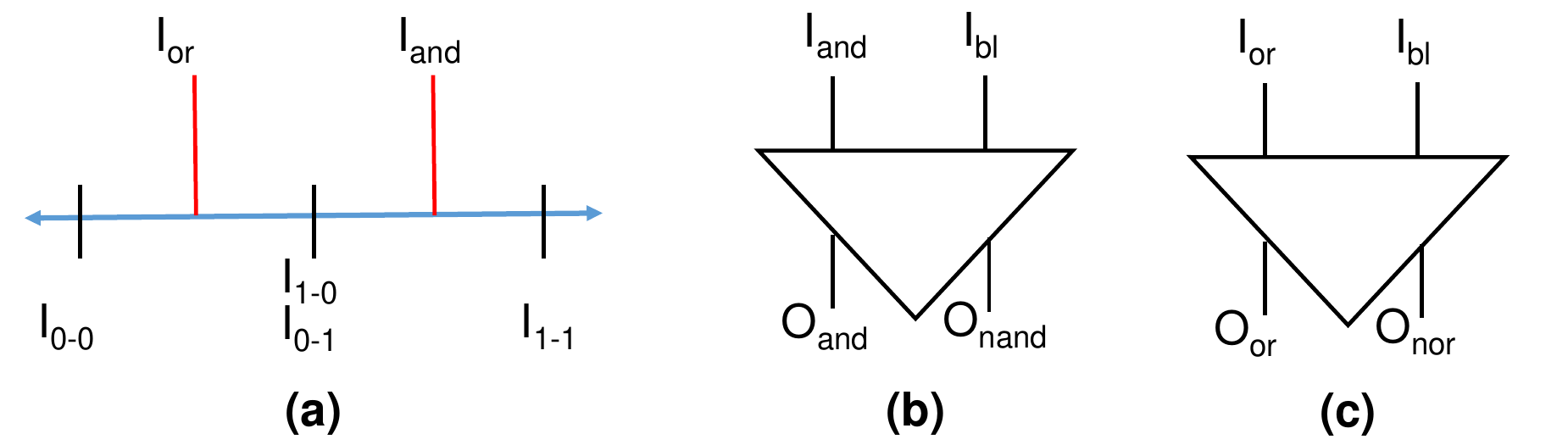}
		\caption{The sensing architecture used in our work (similar to prior work \cite{jain2017computing}), which works for both relaxed retention and non-volatile STT-RAM computing. (a) shows the sensed current for multiple word-lines and the reference signal position; (b) and (c) shows the logical compute circuits.}
		\label{fig:sense}
\vspace{-12pt}
\end{figure}
\begin{figure*}[t]
	
		\centering
		\includegraphics[width=0.55\linewidth]{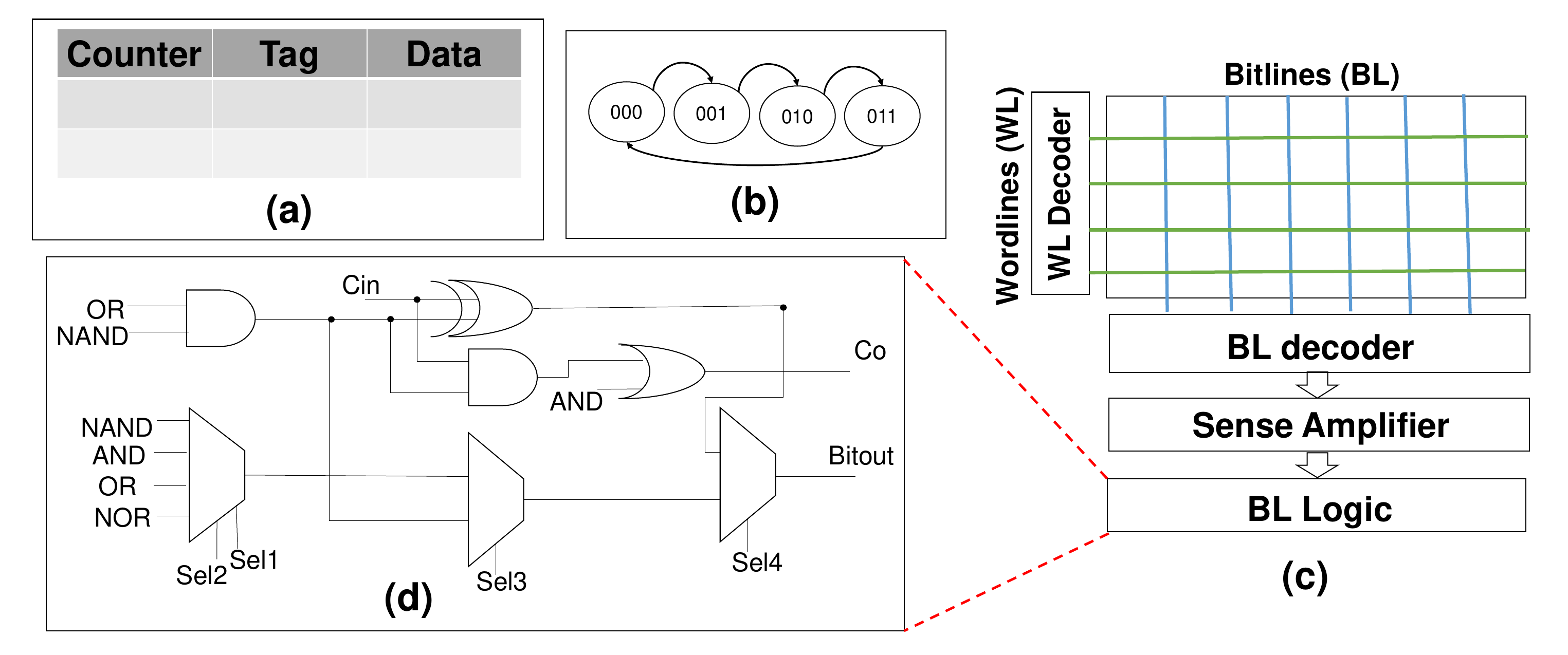}
		\caption{(a) shows the high-level structure of a cache block (b) illustrates the cache block monitor counter implemented using a finite state machine; (c) shows the subarray of a cache with computational logic block after the sense amplifier; and (d) shows the computational logic.}
		\label{fig:architecture}
\vspace{-15pt}
\end{figure*}

This section describes our PiC architecture, retention time selection, optimization for CPU-dependent workloads, and design choices for mitigating process variations.

\subsection{Architecture}\label{sec:arch}

STT-RAM presents significant benefits over SRAM, particularly for PiC applications. For instance, due to its higher write current requirements than read currents, STT-RAM reduces the likelihood of unintended writes when multiple word lines are read for bit-line computing. Consequently, unlike SRAM, multiple word-lines can be activated easily in STT-RAM without corrupting the data, alleviating some complexities associated with SRAM-based computing. For example, there is no need to decrease the cache operating frequency, as required in SRAM systems \cite{jeloka201628}.  

As discussed in Section \ref{sec: background}, bits '0' or '1', stored in the STT-RAM cell, can be represented as resistors with different resistance values \texorpdfstring{R\textsubscript{P}}{} for parallel state and \texorpdfstring{R\textsubscript{AP}}{} for the anti-parallel state, respectively. The difference in \texorpdfstring{R\textsubscript{P}}{} and \texorpdfstring{R\textsubscript{AP}}{} between the stored bits is called the \textit{tunnel magneto-resistance (TMR) ratio}. While sensing, a bit-line will have different current values based on the stored bit. Similarly, sensing the currents for multiple word-lines will result in different output current values, as shown in Figure \ref{fig:sense}a. Based on bits 1 or 0 in these cells, there are three possible current values: \texorpdfstring{I\textsubscript{0-0}}{}, \texorpdfstring{I\textsubscript{1-0}}{}, or \texorpdfstring{I\textsubscript{0-1}}{} and \texorpdfstring{I\textsubscript{1-1}}{}. As such, reference currents can be used to directly compute AND/NAND operations (Figure \ref{fig:sense}b) or NOR/OR operations (Figure \ref{fig:sense}c).  

\vspace{5pt}
\noindent\textbf{PiC architecture.} Figure \ref{fig:architecture} provides a high-level overview of our PiC architecture with relaxed retention. As depicted in Figure \ref{fig:architecture}a, each cache block incorporates a retention time counter to prevent data corruption caused by data expiration. This counter ensures data stability by evicting the block or writing it back to a lower memory level when the retention time is about to expire. We employed an $N$-state finite-state machine (Figure \ref{fig:architecture}b) to implement the counter, which starts in the initial state upon block write, counts up until the retention time is nearing expiration, and then raises a flag. In our study, we assumed $N$ = 4, resulting in a 2-bit overhead per block (Figure \ref{fig:architecture}b), and the counter operates with a clock period of 18.75$\mu$s \cite{kuan2019energy}. The area overhead of the monitor counter is only 0.78\% per block. Designers can increase the counter size $N$ to achieve finer control precision using the same clock input without significantly impacting the cache's area, routing costs, or critical path.

The cache is organized into mats, which are further divided into subarrays. The subarrays in each mat contain a group of word-lines associated with a single sense amplifier, and each word-line is connected to a stored array for parallel computations. The architecture activates two word-lines simultaneously to perform arithmetic or logical operations using the combinational logic illustrated in Figure \ref{fig:architecture}d. The architecture supports addition and logical operations, while the CPU handles more complex operations like multiplication to maintain simplicity in the PiC implementation. Since the implemented operations are associative, they can be executed by sensing multiple word-lines, and the desired operation's output can be selected through multiplexers controlled by the cache controller.

\noindent\textbf{Supporting regular cache operations.} 
Given a PiC-enabled system, performing regular cache operations for CPU-based computing may still be necessary. This paper explores two types of PiC caches: homogeneous STT-RAM caches and heterogeneous STT-RAM caches. The homogeneous design features a uniform retention time throughout the cache, whereas the heterogeneous design is a multi-banked cache with high and low retention time banks. The high retention time cache banks cater to regular cache references from the CPU, and low retention time banks are used for PiC operations. We discuss the heterogeneous cache design in detail in Section \ref{sec:hetdesign}, whereas in this section, we discuss regular cache operations with homogeneous STT-RAM caches that feature the same retention time for both CPU and PiC operations. 

We present two design choices for a homogeneous STT-RAM cache design for regular cache operations. 
The first option involves employing additional sense amplifiers, in addition to the sense amplifiers described in Figure \ref{fig:sense}, to sense bit '0' or '1'. Although this increases the area overhead, it results in lower access latency as the compute elements of PiC operations can be bypassed during CPU cache references. Alternatively, PiC sense amplifiers (Figure \ref{fig:sense}c) can be utilized, which have a reference current tuned to read two cache word-lines: one word-line is read during the cache operation, while the second is set as bit '0'. The sensed output from the sense amplifier is obtained as an \textit{OR} output in PiC-based computation, compared to the reference \texorpdfstring{I\textsubscript{OR}}{}, and read through the multiplexer's OR output (Figure \ref{fig:sense}d).
While this option may reduce the area overhead, it can potentially slow down cache read operations if the compute elements introduce significant latency overhead. In our work, we adopted the latter design choice as the compute elements utilized in this design are simple and do not impose significant latency overhead that would affect the cache access cycles. The write operations for CPU-based computing and PiC utilize the same design circuits since only one word-line is activated during cache block writes or when storing the PiC computation results.

\vspace{5pt}
\noindent\textbf{Scaling the parallel computations.}
To enhance the parallelism of the cache, modifications need to be made to the number of subarrays and the cache geometry. This is a straightforward process for SRAM caches, as the number of subarrays and sense amplifiers can be easily increased \cite{compute_caches}. However, in the case of relaxed retention STT-RAMs, the cache block monitor counters must also be considered, as they are reset during write operations. In PiC, the counter is reset when the computed results are stored in the cache block. If a cache block is partially updated, the unchanged words within the block may expire without resetting the cache block monitor counter. In such scenarios, it is necessary to use a complete cache block to store the computed results in order to avoid discrepancies in block monitor counter updates and prevent data corruption caused by an elapsed retention time. Consequently, the number of compute units in relaxed retention STT-RAM PiC must be a multiple of the cache block size. For instance, let's consider a 64B block cache capable of performing 16 parallel 32-bit integer computations. To increase the number of parallel computations to 32, the cache geometry can be adjusted to increase the number of subarrays, enabling two cache blocks (128B) to be updated simultaneously. This ensures the counter remains simple and seamlessly integrated with relaxed retention caches for CPU and PiC operations.

\subsection{Determining the best retention time}\label{sec:ret_sel}

One of the primary challenges in STT-RAM-based hierarchical in-memory computing (HiMC) is the trade-off between retention time and energy consumption. 
A lower retention time reduces the write energy consumption but may increase the cache miss rates, impacting the overall system performance.
As such, it is imperative that the retention time be sufficient for the cache block lifetimes of the executing workloads. A longer retention time than necessary will incur write overheads. In comparison, a shorter retention time will result in premature expiry of data blocks, leading to high miss rates and data movement overheads. Prior works mapped the retention time to workloads' execution characteristics in traditional CPU-based processing \cite{kuan2019energy, gajaria2019scart}. However, we empirically found that such a scheme is unnecessary for PiC. In traditional computing, the retention time depends on an application's average cache block lifetime and how frequently the data blocks are accessed. For PiC, the cache begins processing as soon as the data is made available in the cache. Thus, the retention time requirement $RT_{req}$ can be expressed as:

\[RT_{req} = (T_p + T_{rp} + T_{mem} + T_{ov})* k, \]

where $T_p$ is the amount of time required to complete the processing of data in the cache, $T_{rp}$ is the amount of time taken for read and write operations, taking into account the reduced retention time, $T_{mem}$ is the time taken to bring the data from the lower-level of memories into the cache, $T_{ov}$ is the time spent for additional overheads, such as error correction, cache management, and other miscellaneous operations, and $k$ is the time taken for a new data load before it can perform PiC operations with the oldest cache block. That is, $RT_{req}$for PiC depends on the time taken to bring the required operands into the cache for computation. For instance, given a computation $c = a + b$, the block containing word $a$ (say, $block A$) only needs to remain in the cache long enough to bring $block B$ (containing word $b$) into the cache to complete the computation. If it takes 100 cycles (or 50ns at 2GHz) to bring $block B$ from memory to cache and 3ns for the $T_p$ and $T_{ov}$, and 2ns for $T_{rp}$ , then for a 32kB cache size, in the worst case, assuming that the first cache block loaded from memory will perform PiC computations with the last loaded cache block ($k$=512), a 28.16$\mu$s retention time is required.

To determine the appropriate retention times, we first analyzed the workloads to determine the average miss latency for each level of the cache hierarchy. Based on this miss-latency, we selected the retention time so data block expiration does not occur during PiC computations.  
Based on our analysis, we found that 75$\mu$s sufficed for the L1 cache and 10ms sufficed for the L2 cache in both PiC and CPU-based computing while minimizing the premature expiry of data blocks. Note that different retention times may be required for different sets of workloads. For example, workloads with high data reuse may require data to be fetched more frequently from the cache than from the main memory. In such cases, a shorter retention time might suffice for PiC. However, the choice of specific retention times is orthogonal to the rest of our analysis. 

\subsection{Mitigating the effects of process variation}

\begin{figure}[t]
	
		\centering
		\vspace{-7pt}
		\includegraphics[width=0.65\linewidth]{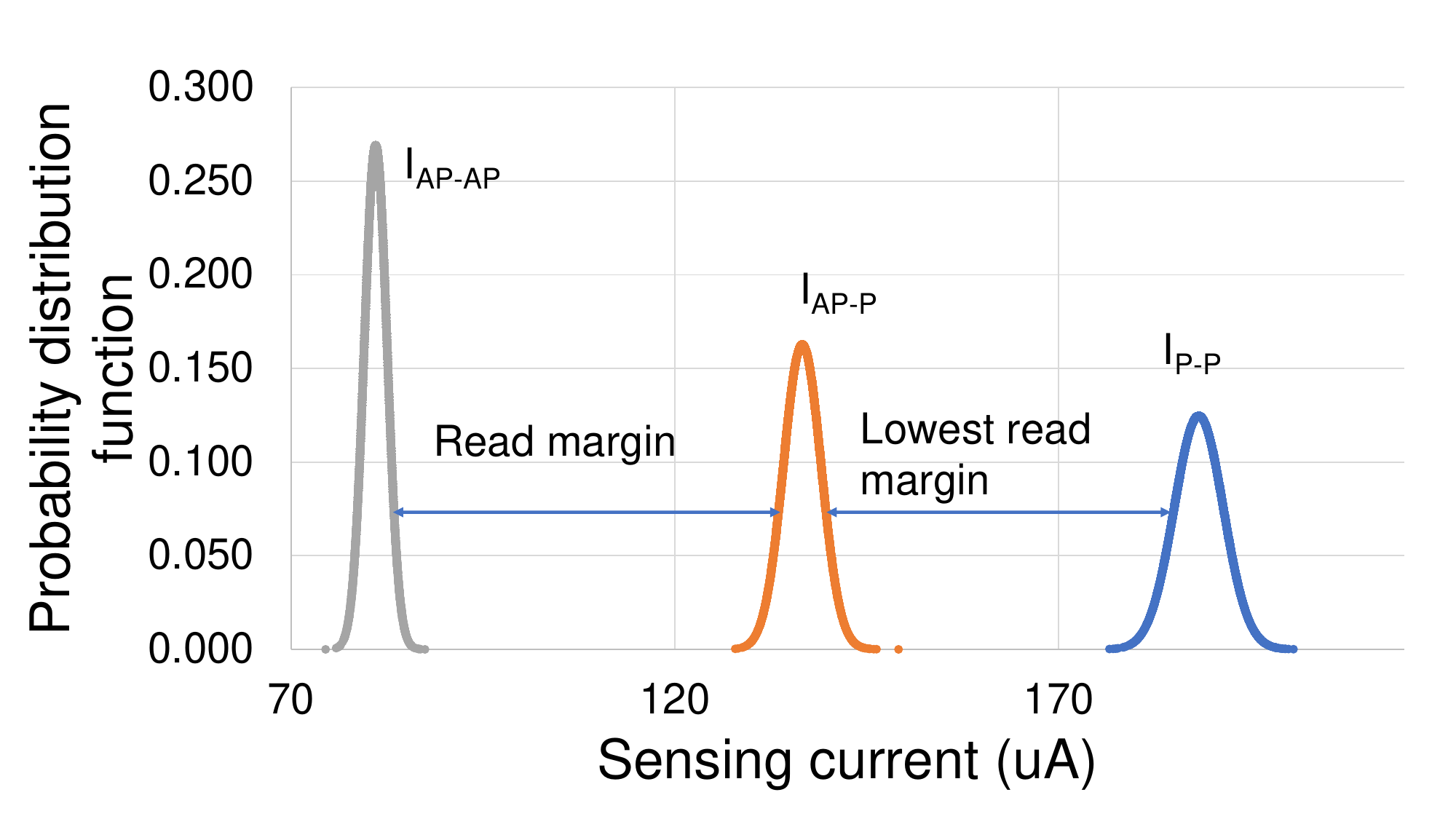}
		\caption{The probability distribution function of the sensing current for PiC STT-RAM under 5\% process variation for 10,000 samples.}
		\label{fig:process_variation}
\vspace{-15pt}
\end{figure}

Another important challenge that arises in the design of STT-RAM HiMC is the susceptibility of reduced retention STT-RAM to process variations, which can lead to higher error rates. Hence, we explored how to mitigate the impacts of process variations in our HiMC architecture.
When conducting a read operation, bits 0 and 1 exhibit different resistance values (TMR ratio), resulting in a change in the output current of the sense amplifier. We simulated multiple retention times by altering the STT-RAM cell parameters, such as the free layer thickness and anisotropy constant ($Hk$), while maintaining a constant TMR ratio. This approach allowed us to utilize the same sense amplifier design for all retention times. For PiC computations, the sense amplifier needs to sense three levels: \texorpdfstring{I\textsubscript{0-0}}{}, \texorpdfstring{I\textsubscript{1-0}}{}, or \texorpdfstring{I\textsubscript{0-1}}{} and \texorpdfstring{I\textsubscript{1-1}}{}, as explained in Section \ref{sec:arch}.
Previous research \cite{jain2017computing} determined that a TMR ratio of 124\% was sufficient for bit-line computing in STT-RAM to ensure reliable read operations across multiple word-lines. However, in our study, we opted for a TMR ratio of 150\% as it resulted in a more distinct difference in the sensed current output between bits 0 and 1, thereby enabling more reliable and distinguishable current levels.

To analyze the impact of process variations, we conducted Monte Carlo simulations \cite{monte-carlo} with 10,000 samples, considering varying STT-RAM cell resistance values using SPICE. Figure \ref{fig:process_variation} depicts our findings on the difference in current levels for STT-RAM PiC under process variations. In our experiments, we set the \texorpdfstring{R\textsubscript{AP}}{} and \texorpdfstring{R\textsubscript{P}}{} under 5\% process variation, following a similar approach as prior studies \cite{jain2017computing}. As observed in the figure, \texorpdfstring{I\textsubscript{AP-AP}}{} passes through the highest resistance, \texorpdfstring{R\textsubscript{AP-AP}}{}, resulting in a low sensing current. \texorpdfstring{I\textsubscript{AP-AP}}{} exhibits a low standard deviation in the sensing current under process variation, leading to a higher probability distribution. Similarly, \texorpdfstring{I\textsubscript{P-P}}{} passes through the lowest resistance, \texorpdfstring{R\textsubscript{P-P}}{}, and demonstrates a high standard deviation in the sensing current, resulting in the lowest probability distribution function. Additionally, we observed that \texorpdfstring{I\textsubscript{AP-AP}}{}, \texorpdfstring{I\textsubscript{AP-P}}{}, \texorpdfstring{I\textsubscript{P-AP}}{}, and \texorpdfstring{I\textsubscript{P-P}}{} remain significantly distinct even under process variations. The read margin between \texorpdfstring{I\textsubscript{AP-P}}{} and \texorpdfstring{I\textsubscript{P-P}}{} is smaller than the read margin between \texorpdfstring{I\textsubscript{AP-P}}{} and \texorpdfstring{I\textsubscript{AP-AP}}{}. We utilized the smallest read margin to adjust the TMR ratio, enhancing the sensing operation's reliability.

However, employing a high TMR ratio can increase the energy required to switch the bit of the STT-RAM cell. Previous studies have utilized relaxed retention time to mitigate the switching energy (write energy) of STT-RAM. When the retention time is relaxed, the switching energy can be adjusted by either reducing the write current or the write pulse \cite{sun2011multi, jog2012cache}. In alignment with prior research \cite{jog2012cache, sun2011multi}, we maintained a constant write current while varying only the write pulse for different retention times. This approach facilitates the establishment of a lower read current than write current for all retention times, thereby reducing read errors \cite{chen2010advances}.

\subsection{Operation chaining for CPU-dependent workloads}\label{sec:chain}

\begin{figure*}[t]
	
		\centering
		\includegraphics[width=0.65\linewidth]{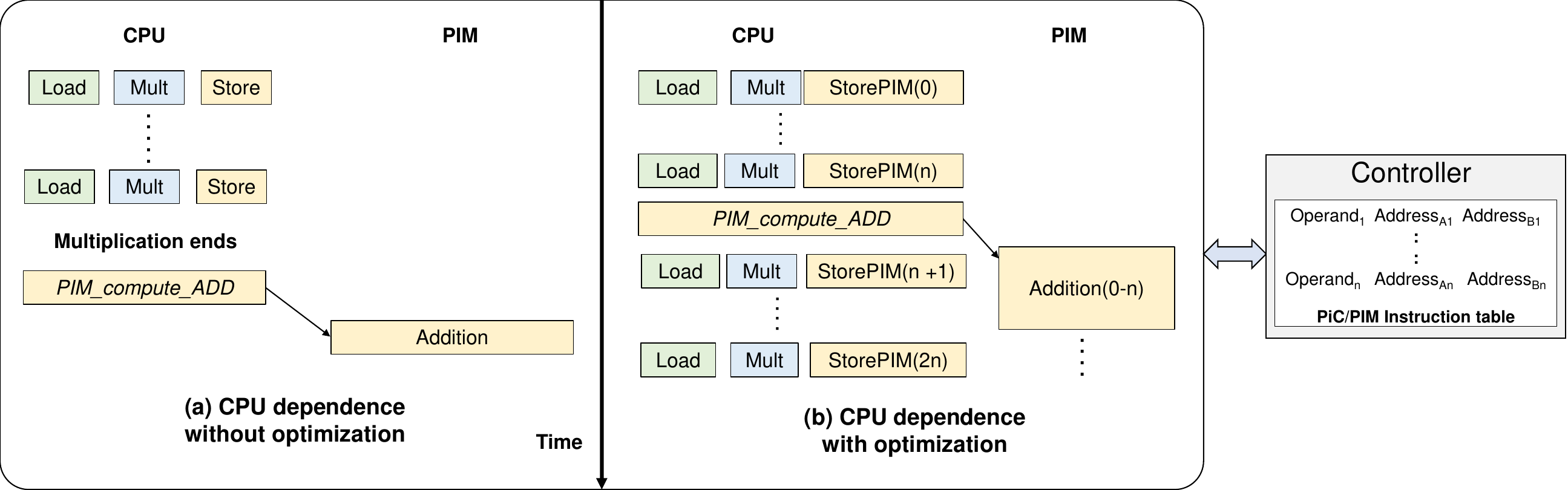}
		\caption{Illustration of the program flow for traditional vs. operation chained PiC/PiM.}
		\label{fig:cpu_dependence}
\vspace{-12pt}
\end{figure*}

Considering the potential overhead associated with CPU-dependent workloads, our objective was to enhance parallelism in order to minimize the waiting time for CPU results. Previous research \cite{boroumand2018google} has demonstrated that traditional PiM divides the program into PiM and CPU execution segments, with PiM commencing only after CPU execution is complete. This computational model is illustrated in Figure \ref{fig:cpu_dependence}. Our workload analysis identified the potential for optimizing CPU-dependent workloads to exploit better parallel processing between the PiC/PiM and CPU execution units. This optimization, referred to as \textit{operation chaining}, draws inspiration from architectures employing \textit{vector chaining} \cite{dongarra1984implementing}. As depicted in Figure \ref{fig:cpu_dependence}, operation chaining enables the utilization of interim CPU results for PiC/PiM computations without introducing additional memory references. After each computation, the processor stores the intermediate results in memory or cache for subsequent PiC/PiM computations.

As seen in the figure, the data is loaded in the CPU for computations and the data is then stored in the respective PiC/PiM location using the \emph{StorePIM(n)}. The data address is also recorded in the memory/cache controller and is stored in an instruction table. 
The CPU communicates via a PiC/PiM controller using a \emph{Compute} signal to start the PiC/PiM computations and receives a \emph{DONE} signal indicating that the computations have been completed. During the \emph{Compute} signal, the PiC compute instruction is sent to the cache controller. The specific PiC/PiM instruction starts executing on the addresses recorded in the instruction table.
When a \emph{Compute} signal is received, the controller checks to see if the instructions are suitable for bit-line computing and start performing computations. More information on the cache controller management can be found in Section \ref{sec:het_cache_management}.


The size of the instruction table depends on how many addresses are stored before a \emph{Compute} instruction is issued. A larger instruction table will have higher overheads, but a smaller instruction table might result in low utilization of compute units, leaving some idle. Thus, an ideal instruction table size should be sufficient to ensure that all the compute units are utilized while executing the \emph{Compute} instruction. For example, given a memory hierarchy with 16 32-bit compute units, assuming 32-bit block addresses, the memory overhead will be 128B.

Operation chaining is enabled by the compiler, which detects the data dependencies and ensures the absence of potential hazards. The programs are augmented with new instructions: \textit{StorePIM\_MEM} (to store data in the appropriate PiC/PiM location specified by \textit{MEM}) and \textit{Compute\_Inst\_PIM\_MEM} (to initiate execution in the PiC/PiM architecture in the respective memory hierarchy). 
Thus, using operation chaining, the execution latency of the PiC/PiM is effectively hidden behind the processor's execution latency.

\section{Heterogeneous cache design}\label{sec:hetdesign}

To optimize both CPU-based computing and PiC, in this section, we explore a heterogeneous cache design that features multi-retention time STT-RAM caches. The heterogeneous cache consists of a high-retention time STT-RAM region for CPU-based computing and a lower retention time for PiC computing. A lower retention time region can reduce the write overheads whereas a higher retention time will increase the write overheads but will have fewer cache misses due to cache blocks expiring \cite{kuan2019energy,gajaria2019scart}. For CPU-based computing, the cache block lifetime is much longer than in PiC-based computing, since blocks must be held in the cache for much longer. For PiC-based computing, on the other hand, intermediate results are stored in the same level of the memory hierarchy, resulting in shorter block lifetimes. Moreover, highly parallel execution of the program also reduces the reuse distance of the data blocks, thus requiring a lower retention time. 

A heterogeneous cache design can be targeted for CPU-dependent workloads. However, it also can be used for other CPU-independent workloads if the program initialization will be handled by the CPU i.e. the CPU initializes the program and keeps track of the program counter, and PiC/PiM execution acts as an accelerator that executes computations initiated by the CPUs. In this section, we propose a heterogeneous cache architecture and present a low-overhead heterogeneous cache management technique. 

\subsection{Heterogeneous retention cache architecture}
\begin{figure}[t]
		\centering
		\includegraphics[width=\linewidth]{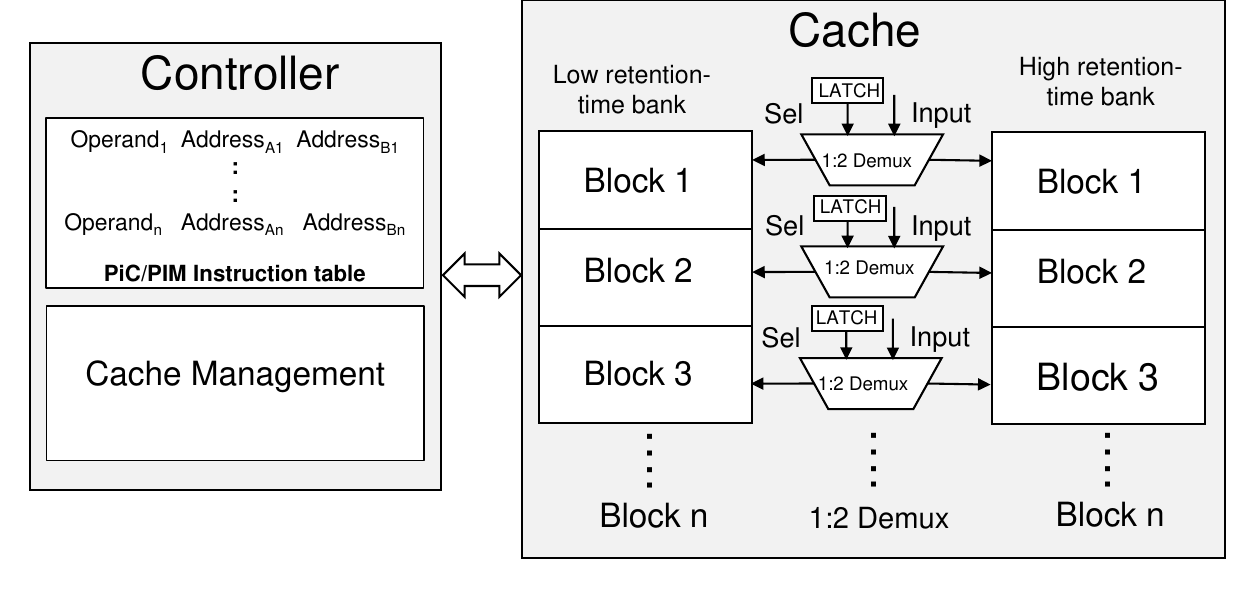}
		\caption{Heterogeneous retention time architecture with low and high retention time cache blocks. The architecture has a latch and a 1:2 demultiplexer for each cache block that determines the bank to be used to access the specific cache block.}
		\label{fig:het_arch}
\vspace{-15pt}
\end{figure}

\begin{figure}[ht]
\centering

    \begin{subfigure}[t]{\linewidth}
      \centering
      \includegraphics[width=0.95\linewidth]{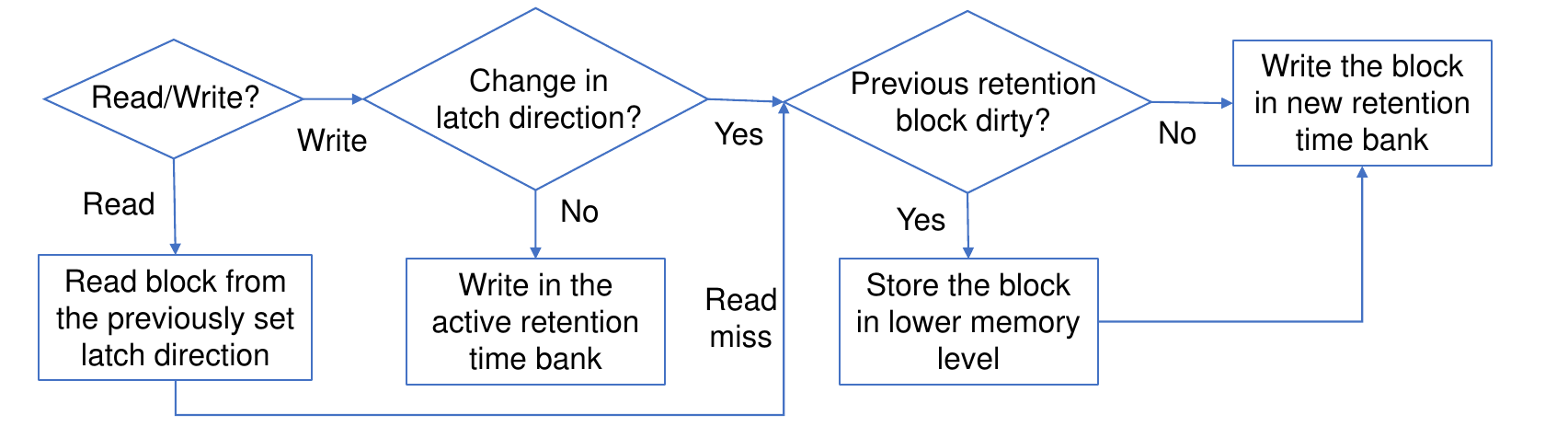}
      \caption{Latch control}
      \label{fig:lc}
    \end{subfigure}
    \begin{subfigure}[t]{\linewidth}
      \centering
      \includegraphics[width=0.95\linewidth]{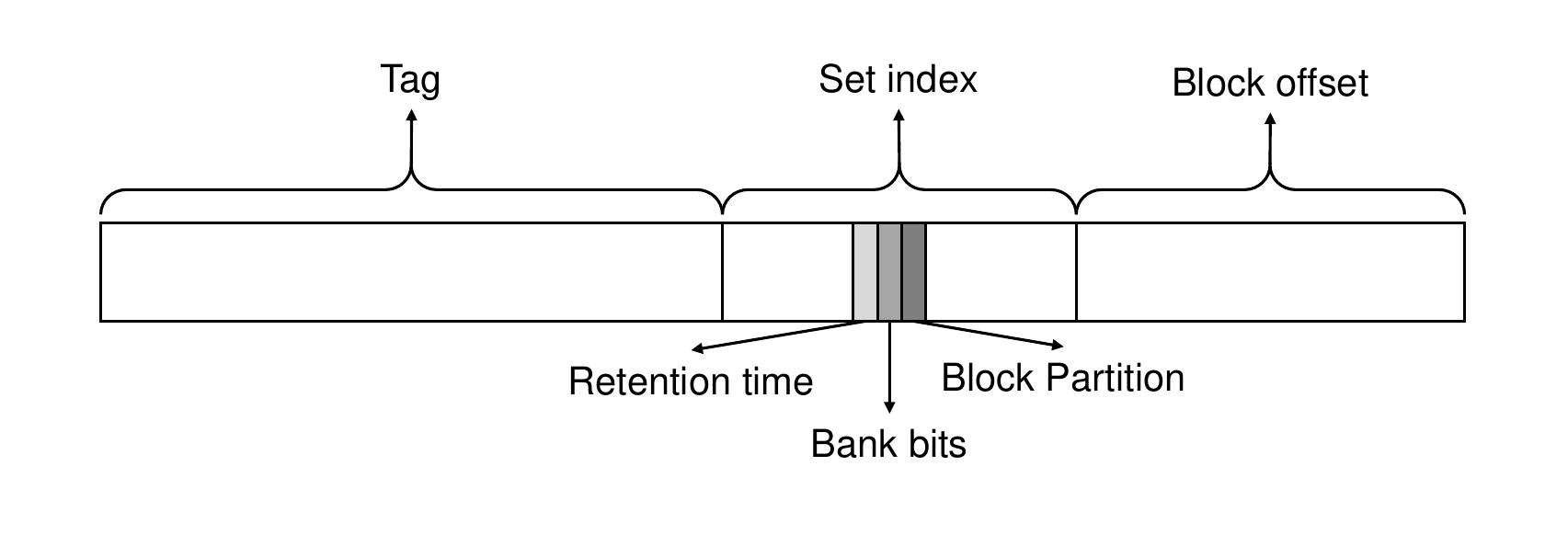}
      \vspace{-7pt}
      \caption{Address/bits information. The design of the bank bits and block partition bits used in our work is similar to prior work \cite{compute_caches}.}
      \label{fig:bits}
    \end{subfigure}%
\caption{Latch control and cache address information.}
\label{fig:latch_control}
\vspace{-15pt}
\end{figure}

Heterogeneity in PiC and PiM can be achieved using two cache designs. One design involves different retention time cache banks, and the other involves subarrays of different retention times, in which the subarrays share the H-tree and routing wires. We find that heterogeneity at the bank level results in a higher area overhead, as the routing resources within the cache bank are not shared. On the other hand, heterogeneity at the subarrays results in a higher routing overhead cost for the wires in the cache bank. It thus increases the overall access latency of the cache, impacting the performance of CPU-based computing as well. However, since STT-RAMs are denser than SRAMs (by a factor of 3 to 9) \cite{kuan2019energy}, and we aim to minimize the latency overhead, we consider the area overhead from bank-level heterogeneity to be acceptable for this implementation. Despite the area overhead, our approach still reduces the overall area compared to SRAM implementations. Therefore, we employ heterogeneity at the bank level for our work.

Our approach allows a switch between the different retention time regions, depending on whether PiC- or CPU-based computing is being performed. Figure \ref{fig:het_arch} illustrates the architecture of the heterogeneous STT-RAM cache. The architecture comprises two banks of retention time, in which each cache block has its own latch along with a 1:2 demultiplexer, which determines the retention-time region when a cache block is accessed. At any given time, the data in a cache block is only present in one retention time bank. The latch is a one-bit memory that controls the select line of the demultiplexer and thus selects which retention time bank to use for cache data access. The decision on the retention time of the cache block is driven by whether the code is to be run using the CPU or PiC/PiM architectures, which we discuss in Section \ref{sec:het_cache_management}. Note that this configuration, with dual cache banks, does not result in high static power consumption due to STT-RAM's near-zero leakage costs. We discuss the impact of having multiple banks active on the overall energy consumption in detail in Section \ref{sec:comp_sram_sttram}.

\subsection{Heterogeneous retention cache management}\label{sec:het_cache_management}
In this section, we present the heterogeneous retention cache management strategy. For efficient cache operation, we leverage the fact that the lower-retention time bank will remain active during PiC computing, and the high retention time bank will be during CPU-based computing. Therefore, we only change the latch bit when a cache block is written. Figure \ref{fig:lc} shows the latch management. During the read operation, the cache block is read from the retention time bank according to the latch bit. During the write operation, the latch direction is updated on the basis of the retention time information that serves as input to the latch. The latch direction can be either updated by the incoming address in case of \textit{StorePIM} instruction as seen in Section \ref{sec:chain} or by the cache controller to bring the data from other levels of the memory. 

The compiler can help greatly in determining the retention time selection. For example, during the write operations, the data can be automatically placed into the low-retention time banks during operation chaining. The compiler can identify whether the next set of instructions will be performed by the CPU or by PiC compute units, as discussed in Section \ref{sec:chain}. Thus, the data going into the cache for PiC computations can be identified and directly placed into the low-retention time banks. Moreover, if the data are not present in the cache in case of a read miss, it can be fetched from the lower levels of the memory and placed in the appropriate cache bank based on whether the instruction is executed on the CPU or on PiC and the latch direction is set accordingly. However, during the read operation, if the data are already present in the cache, regardless of the location of the bank, it can be fetched by looking at the latch bit. Thus, a cache block would only switch banks if a write operation occurs and thereby greatly reduces the data transfer overhead between the low and high retention time banks.

To facilitate the latch mechanism, we use the address bits of the cache block which includes the retention time bit, bank bits, and block partition. The retention time bit indicates whether a low or a high retention time is needed and will only be checked whenever a write operation (store or cache miss) occurs in the cache. Whenever there is a cache write, the retention bit is overwritten in the latch for the respective cache block. Bank bits and block partitions operate similarly to prior work \cite{compute_caches}. The bank bits indicate the bank location and the block partition indicates the subarray location for a given bank for which the data to be computed are bit-line aligned. Compiler modifications can be used to ensure that the data are bit-line aligned, similar to prior works. For example, Fujiki et.al. \cite{fujiki2019duality} proposed a cache architecture, called \emph{Duality Cache}, that adopts a single instruction multiple threads (SIMT) execution model where the cache architecture acts as both vector processing units and register file, and schedules VLIW instructions for in-cache operations. The authors adopt the CUDA/OpenACC compiler model to translate instructions to their in-house \emph{Duality cache ISA} to ensure bit-line alignment.  Wang et. al. \cite{wang2022infinity} proposed using a novel execution and intermediate representation compiler model that automatically orchestrates data management and performs runtime layout transformations for bit-serial execution. While both works ensure bit-line alignment, we adopt the approach by Wang et. al. \cite{wang2022infinity} work for our STT-RAM-based HiMC due to higher flexibility and lack of architectural assumptions required for the \emph{duality cache}, such as CUDA compiler, which are unavailable for efficient in-memory compute operations.

The overall execution of the heterogeneous retention PiC system is given by Algorithm \ref{algorithm:management}. At program initialization, the latches default to pointing to a high retention time bank to enable CPU computing during the initialization. Then, when PiC-based instructions are ready to be executed, they are added to the PiC instruction table maintained by the cache controller (line 2). The instructions contain the operand and the address of the entire block. The number of compute units is equivalent to the block size (lines 6-7), which ensures that the data on which PiC/PiM computing occurs is bit-line aligned, provided that the data blocks are located in the same retention time zone, bank, and subarray, as discussed in Section \ref{sec:arch}. 

Before executing a particular PiC instruction, the cache controller checks whether the data block satisfies the bit-line alignment criteria described above. This step is important when PiC computations have to be done on data that was previously stored by the CPU in the high-retention-time region. If the criteria are not met for a PiC instruction, the controller moves data present in the high retention time bank to the low retention time bank and places it in the same bank and subarray (line 11). This is done by issuing a write operation to the other retention time bank and also by setting the \textit{change in latch direction flag} in the latch control (in Figure \ref{fig:lc}) to \emph{TRUE}. If the data is not present in the cache due to a read miss, the cache controller fetches it from other memory levels and places it in the appropriate cache bank based on whether the instruction is executed on the CPU or PiC. Until then, the controller checks if other sets of instructions stored in the PiC/PiM instruction table are bit-line aligned and executable (line 7) to reduce the stalls caused by the data movement. However, this requires the availability of more instructions in the instruction table, which will increase the instruction table size. The instruction table size can depend on the number of parallel compute units present in the memory, as discussed in Section \ref{sec:chain}. 

\begin{algorithm}[ht]
\caption{Cache controller management}
\label{algorithm:management}
\SetAlgoLined
    \SetKwInOut{Input}{Data}
    \SetKwInOut{Output}{Result}
    \DontPrintSemicolon
    \Input{
    List of instructions to be executed in PiC, 
    
    $T = [t_1, t_2, ..., t_n]$\
    
List of latch state table, 

$L = [l_1, l_2, ..., l_m]$, where $l_i \in {0,1}$ and $m$ is the number of cache blocks    
    }
    \Output{Heterogeneous cache PiC execution}

$\forall$ L = high retention time



\For{instructions $\in$ program}
{

$(\text{RetentionTime, Bank}) _{\text{A, B}} \longleftarrow \text{Address} _{\text{A, B}}$

$(\text{Read/Write}) _{\text{A, B}} \longleftarrow \text{Instruction} _{\text{A, B}}$

\emph{LatchControl(Read/write)}

\If{instructions = PiC}{
\For{each PiC instruction}{

\If{
$\text{RetentionTime}_{\text{A}} = \text{RetentionTime}_{\text{B}} \textbf{ and } \text{Bank}_{\text{A}} = \text{Bank}_{\text{B}}$ }
{
Execute the instruction.
}
\Else{
 Transfer the cache block to a low retention time bank.
 
 \If{
 $\text{CacheMiss}_{\text{A}} \textbf{ or } \text{CacheMiss}_{\text{B}}$}
 {
Fetch the data from the lower level of memory. 

Move to other PiC instructions to avoid stalls. 
 
 }
}
}
}
}

\end{algorithm}

\section{Experiments}\label{sec:exp}

In this section, we first discuss the workloads that we used for our analysis. Then, we describe our experimental setup for the energy and latency access for STT-RAM memory, STT-RAM, and SRAM caches. 

\subsection{Workloads}

\begin{table}
\caption{Workloads used in our experiments}
\footnotesize
\centering
\scalebox{0.85}{
\begin{tabular}{|l|c|c|}
\hline
Category & Kernel & Input size  \\ \hline
\multirow{4}{*}{CPU-dependent} & Kmeans nearest neighbor (\textit{KNN})& 10$^5$ nodes\\
& 2D convolution (\textit{conv}) & 10$^6$ samples \\
& Histogram (\textit{hist}) & 10$^6$ samples \\
& Root-mean square error (\textit{rmse}) & 10$^6$ samples \\
\hline
\multirow{3}{*}{\parbox{2.2cm}{CPU-independent,\\ high data reuse}} & Binarized neural network (\textit{bnn}) & 10$^6$ samples\\
& Matrix addition (\textit{mat\_add}) & 10$^6$ samples\\ 
& String Comparison (\textit{string}) & 2,409,780 letters\\  
\hline
\multirow{2}{*}{\parbox{2.2cm}{CPU-independent,\\ low data reuse}} & Carryless multiplication (\textit{cmul}) & 10$^6$ samples \\
& & \\
\hline
\end{tabular}}
\label{tab:workload}
\end{table}
To simulate the behavior of real-world applications, we utilized a set of eight workloads with varying characteristics. The workloads and their corresponding input sizes are presented in Table \ref{tab:workload}. For our analysis, we classified the workloads into three distinct groups. The first group consists of \textit{CPU-dependent workloads}, which predominantly exhibit characteristics that make them more efficiently executed by the CPU. These characteristics include low instruction-level parallelism, operations involving pointers, complex branching conditions, or computationally intensive tasks that are not well-suited for in-memory computing designs (e.g., multiplication operations in our study). The second group comprises \textit{CPU-independent workloads}, which are simpler kernels that can be executed entirely using PiC/PiM. Previous research on in-memory computing has primarily focused on these types of workloads \cite{compute_caches, parveen2018hielm}. Within the CPU-independent workload group, we further categorized them based on high data reuse and low data reuse, allowing us to evaluate the trade-offs associated with data movement overheads for relaxed retention STT-RAM PiC and non-volatile STT-RAM PiM. These selected workloads are commonly encountered in applications such as image/signal processing, data querying, etc.

\subsection{Experimental methodology}
\begin{table*}[t]

\renewcommand{\arraystretch}{0.95}
\caption{Cache and memory configurations}
\label{tab:cycles}
\centering
\scalebox{0.85}{
\begin{tabular}{|c||cc|ccc|c|}

    \hline
    Memory hierarchy				&\multicolumn{2}{c|}{L1 cache 32kB-64B-4}  &\multicolumn{3}{c|}{L2 cache 1MB-64B-8} & Memory 512MB size\\
    \hline
   Memory Device &SRAM &STT-RAM &SRAM &\multicolumn{2}{c|}{STT-RAM} &STT-RAM \\
   \hline
   Retention time&-- &75$\mu$s &-- &75$\mu$s &10ms &5years \\
    \hline
    Read latency (cycles) &1 &1 &2 &2 &2 &32 \\
    \hline
    Write latency (cycles) &1 &2 &2 &3 &4 &56 \\
    \hline
   Logical operation (cycles) &3 &3 &4 &5 &6 &88 \\
    \hline
    Add operation (cycles) &18 &15 &19 &15 &16 &97 \\
    \hline
    Read energy per bit (in pJ) &0.125 &0.086 &1.77 &0.75 &0.75 &24.55 \\
    \hline
    Write energy per bit (in pJ) &0.19
 &4.69 &0.62 &9.647 &15.604 &640.89\\
    \hline
    Logical computation energy per bit (in pJ) &0.915 &5.376 &2.997 &10.997 &16.954 &666.045 \\
    \hline
    Add computation energy per bit (in pJ) &1.355 &5.816 &3.437 &11.437 &17.394 &666.49
 \\
    \hline
    Leakage power (mW) &43.95 &17.63  &1168.95 &182.8 &182.2 &222.36 \\
    \hline
\end{tabular}}
\vspace{-15pt}

\end{table*}

The study of bit-line computing is quite nascent (currently in the exploration phase), and thus, it is very difficult to come up with a real hardware testbed to study the benefits of PiC/PiM architectures using STT-RAM. Therefore, we have used state-of-the-art simulation infrastructure, just like prior works (e.g., \cite{jain2017computing, compute_caches}), to study the benefits of PiC/PiM computing.
To model the PiC/PiM computation logic, we used SPICE simulations with 22nm CMOS libraries. We used NVSim \cite{dong2012nvsim} to obtain the STT-RAM/SRAM cache and STT-RAM memory access latency and energy. The NVSIM simulator presents a robust and efficient way to evaluate STT-RAM cache and memories since it has been validated against real STT-RAM cells. We modified gem5 \cite{binkert2011gem5} to implement relaxed retention STT-RAM caches and to model SRAM caches. The gem5 statistics were then integrated with McPAT \cite{li2009mcpat} to obtain the total system power. We modeled a processor like ARM Cortex A72 with a 2GHz clock and 8GB memory, of which 512MB is used for PiM. 

Table \ref{tab:cycles} presents the cache and memory configurations, along with the read, write, and computation latencies for each operation at various levels of the memory hierarchy. Logical operations are completed in a single cycle, as they require approximately 120ps to execute bitwise logical operations. To optimize area and energy for ADD operations, we employed a Ripple Carry Adder design, which performs one bit computation at a time and propagates the carry bit to higher significant bits. STT-RAM requires fewer cycles than SRAM for ADD operations due to the longer slack within each STT-RAM read cycle, allowing for more bit operations per cycle.
The energy figures are calculated by considering the energy required to access the subarrays, bit-lines, and memory cells. The energy for logical and ADD computations includes the access energy for the cache, the computation energy, and the energy needed to store the results in a subarray.
Consistent with prior work \cite{compute_caches}, we assumed that the execution unit of the processor is powered off during PiC/PiM to conserve power.

\section{Results and Analysis}\label{sec:results}

In this section, we first compare operation chaining to previous PiC/PiM computing approaches and then analyze the tradeoffs of PiC using STT-RAM vs. SRAM. Subsequently, we contrast STT-RAM-based PiC with PiM and discuss the associated overheads. Additionally, we present results comparing heterogeneous and homogeneous STT-RAM caches for PiC. Lastly, we examine the overheads of PiC and PiM to understand the tradeoffs of the two approaches. 

The results of STT-RAM vs. SRAM and PiC vs. PiM are normalized to a traditional CPU with STT-RAM caches similar to \cite{sun13_stt}. The traditional CPU does not have any PiC/PiM optimization but features a relaxed retention STT-RAM cache of 75$\mu$s at L1 and 10ms at L2. Many prior works have indicated several energy and area benefits of replacing SRAM caches with STT-RAMs in different memory levels of traditional CPU-based computing \cite{kuan2019halls,gajaria2019scart,sun2011multi,sun13_stt}. Thus, using STT-RAM caches with the base CPU allows us to robustly evaluate the added benefits of PiC/PiM architecture using STT-RAMs.
Comparisons are performed with respect to execution latency, energy consumption, and area. It is important to note that the presented results account for the system components remaining active during PiC-based computing, reflecting the overall system energy.

\subsection{Comparison between operation chained and conventional PiC/PiM (prior work)}
 \begin{figure}[t]
		\centering
		\includegraphics[width=0.85\linewidth]{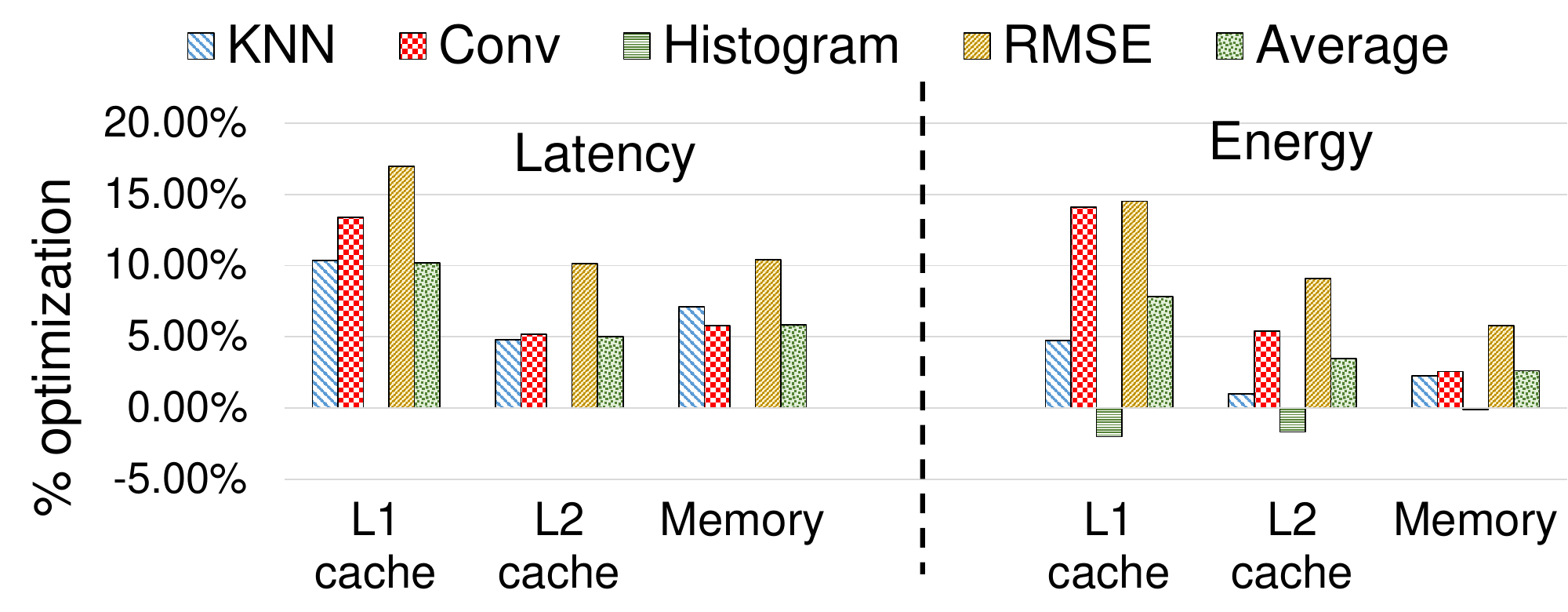}
		\caption{Latency and energy savings of CPU-dependent workloads using operation chaining for PiC/PiM at various memory hierarchy compared to prior work (no operation chaining).}
		\label{fig:opt vs non-opt}
\vspace{-15pt}
\end{figure}

Firstly, we compare the optimized method for CPU-dependent workloads, which utilizes operation chaining (Figure \ref{fig:cpu_dependence}), with the traditional PiM approach as done in previous studies \cite{boroumand2018google}. This comparison is conducted at different levels of the memory hierarchy, employing STT-RAMs for the cache hierarchy (L1 and L2) and memory. The L2 cache is a homogeneous cache design with a long retention time of 10ms. We assume that the processor can operate concurrently with PiC/PiM elements to fully leverage the advantages of operation chaining. Although PiC has previously been implemented exclusively in SRAM \cite{compute_caches,eckert2018neural}, we utilized an STT-RAM PiC implementation to assess the benefits of operation chaining. In contrast to conventional PiC/PiM computing, the execution unit of the processor remains active while utilizing operation chaining.

Figure \ref{fig:opt vs non-opt} demonstrates the comparisons of latency and energy savings between operation-chained and conventional PiC/PiM for L1 cache, L2 cache, and memory. On average, operation chaining improved overall latency for CPU-dependent workloads in L1 cache, L2 cache, and memory by 10.19\%, 5.02\%, and 5.82\%, respectively. Conventional PiC involved substantial data write-backs to lower memory levels, necessitating cache reloads, whereas operation chaining reduced data movement by up to 10.21\% and increased the utilization of compute units. The most significant latency enhancements were observed for $RMSE$ (16.98\%, 10.13\%, and 10.43\% for L1, L2, and memory, respectively), which involved multiple data transfers to the processor for complex computations (e.g., square and square root operations). In the worst-case scenario, operation chaining provided minimal savings for $histogram$ as the majority of the application was executed on the CPU due to complex or sequential operations, resulting in savings of 0.02\%, 0.005\%, and 0.01\% for L1, L2, and memory, respectively.

Operation chaining reduced energy consumption by 7.83\%, 3.45\%, and 2.62\% for L1 cache, L2 cache, and memory, respectively, compared to traditional PiC/PiM. Similar to latency, the most substantial energy improvement was observed for $RMSE$ at 14.52\%, 9.11\%, and 5.78\%, respectively. However, operation chaining slightly increased the energy consumption for $histogram$ by 2.02\%, 1.68\%, and 0.11\%, respectively. This increase in energy was due to the simultaneous activity of the CPU and PiC/PiM execution units, and the latency savings did not fully compensate for the additional power consumption of the PiC/PiM execution units.
Given the overall superiority of the operation-chained PiC/PiM over the traditional PiC/PiM approach, we subsequently perform an analysis of computing across the memory hierarchy using the operation-chained PiC/PiM.

\subsection{Heterogeneous vs. homogeneous STT-RAM PiC architectures}

For our work, we do not use heterogeneous configurations for L1 caches. As seen in Section \ref{sec:hetdesign}, using heterogeneous configurations allows us to reduce the retention time for PiC-based operations thereby reducing the write cycles and energy, and reducing write overheads. 
However, we empirically found that for L1 caches, reducing the write cycles from 2 (for 75$\mu$s as seen in Table \ref{tab:cycles}) to 1 requires a retention time $<1\mu$s, which significantly increases the miss rates for PiC-based computing, resulting in high latency and energy costs.
Therefore, the heterogeneous design is only used in the L2 cache. We performed a design space exploration of average cache block lifetimes as seen in Section \ref{sec:ret_sel} to determine that 75$\mu$s sufficed for the low-retention time bank for PiC-based computing and 10 ms sufficed for the high-retention time bank for CPU accesses. The homogeneous STT-RAM L2 cache will have 10ms of retention time. 

\begin{figure}[t]
		\centering
		\includegraphics[width=0.8\linewidth]{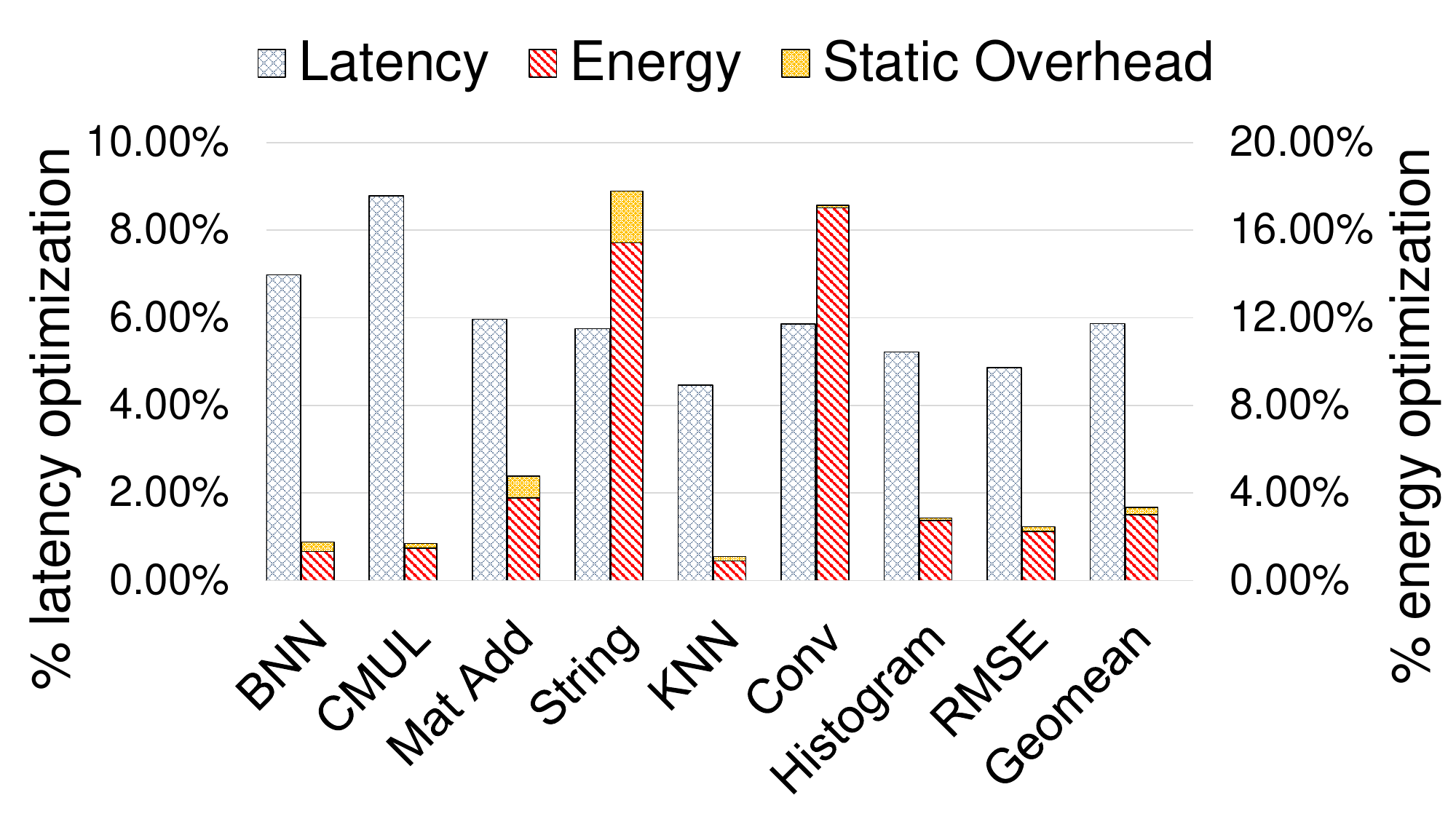}
		\vspace{-7pt}
		\caption{Latency and energy savings for STT-RAM$_{het}$ at the L2 cache for all the workloads compared to a homogeneous STT-RAM cache design with 10ms retention time (STT-RAM$_{10ms}$).}
		\label{fig:het_vs_hom}
\vspace{-10pt}
\end{figure}

Figure \ref{fig:het_vs_hom} presents the latency and energy savings of STT-RAM$_{het}$ compared to STT-RAM$_{10ms}$. The figure also presents the percentage of static energy overhead incurred by the extra active bank (keeping both high and low retention time bank on) for STT-RAM$_{het}$ as seen in Section \ref{sec:hetdesign}. For brevity, the results only show the optimizations achieved for PiC computations since both STT-RAM$_{het}$ and STT-RAM$_{10ms}$ execute the CPU-based operations using 10ms retention time. As seen from the figure, STT-RAM$_{het}$ achieved latency and energy optimizations for all the workloads. The average latency improvement across all the workloads was 5.86\% and up to 8.78\% for $CMUL$. The latency benefits result from lower write latencies achieved by reducing the data movement and shorter compute cycles for 75$\mu$s retention time, as seen in Table \ref{tab:cycles}. The highest latency improvement was observed for $CMUL$ because it comprises entirely of logical operations, and the STT-RAM$_{het}$ reduced the latency of logical operations from 6 to 5 cycles. On the other hand, the other workloads also feature add operations for which STT-RAM$_{het}$ reduced the latency from 16 to 15. As a result, the overall benefits are less for workloads featuring add operations.

For energy optimization, STT-RAM$_{het}$ only achieved average savings of 2. 66\%, with savings of up to 16. 88\% for $Conv$. The highest energy savings were observed for workloads that exhibited high data reuse because they stored more intermediate results in the cache without incurring cache misses. The STT-RAM$_{het}$ architecture is especially suited for these workloads due to a 38.46\% reduction in write energy per bit over STT-RAM$_{10ms}$ as seen in Table \ref{tab:cycles}.
For example, the highest energy savings were observed for $String$, and $Conv$ at 13.07\% and 16.88\%, respectively. $String$ compares new data with previously stored results, while $Conv$ stores intermediate results and performs accumulation operations on these results. 
Moreover, the power consumption of the extra bank did not result in significant energy overhead for STT-RAM$_{het}$.
On average, the power overhead was only 0.34\% across all the workloads. The highest power overhead was 2.36\% for $String$ which had high data movements that increased the latency and static energy overhead of the extra active bank of the heterogeneous architecture.
Despite this static energy overhead, STT-RAM$_{het}$ still resulted in higher energy savings, as seen in Figure \ref{fig:het_vs_hom} for all the workloads. These results reveal the superiority of STT-RAM for PiC and also indicate that a shorter retention time in the L2 cache suffices for PiC, unlike CPU-based computing, which requires longer retention times in the L2 cache.
Therefore, considering the optimization potential of the heterogeneous STT-RAM cache, we use STT-RAM$_{het}$ for L2 caches for the rest of our experiments.

\subsection{Comparison of different PiC candidates}\label{sec:comp_sram_sttram}

In this subsection, we compare different candidates for PiC, including relaxed retention STT-RAM, an SRAM design used in prior work with a low-voltage word-line (SRAM$_{lv}$) \cite{jeloka201628}, and a theoretical ideal SRAM cache model (SRAM$_{ideal}$) that achieves the least data corruption during bit-line computing. SRAM$_{lv}$ has 50\% more delay than SRAM$_{ideal}$ and requires 20\% lower dynamic energy for its operations. The comparisons between SRAM and STT-RAM are presented for L1 and L2 caches, with a homogeneous L1 and heterogeneous L2 STT-RAM. The results presented are normalized to the traditional CPU with STT-RAM caches with retention time of 75$\mu$s at L1 and 10ms at L2 cache.

\begin{figure*}[ht]
\centering
\begin{subfigure}{.5\linewidth}
  \centering
  \includegraphics[width=0.75\linewidth]{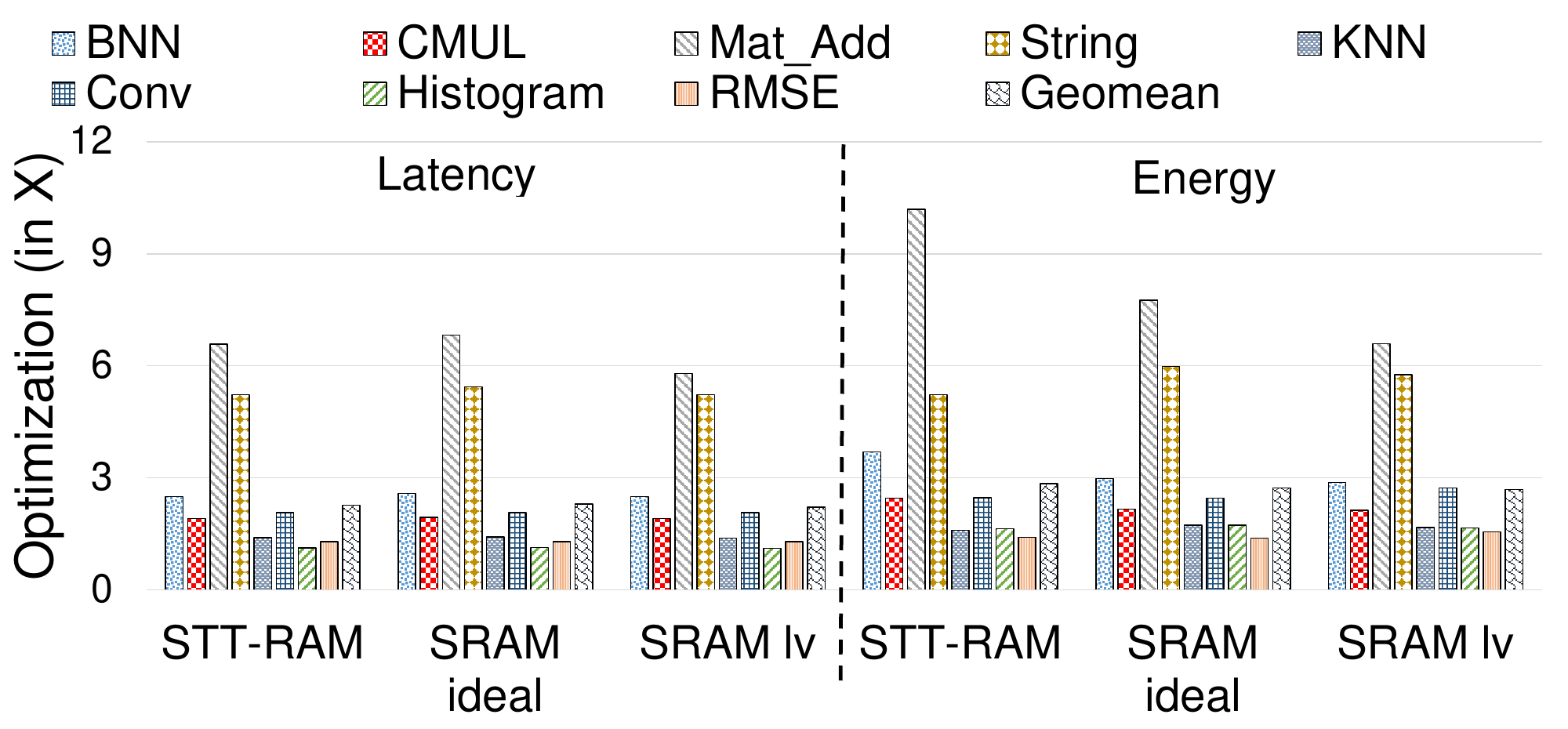}
  \vspace{-10pt}
  \caption{L1 cache}
  \label{fig:l1}
\end{subfigure}%
\begin{subfigure}{.5\linewidth}
  \centering
  \includegraphics[width=0.75\linewidth]{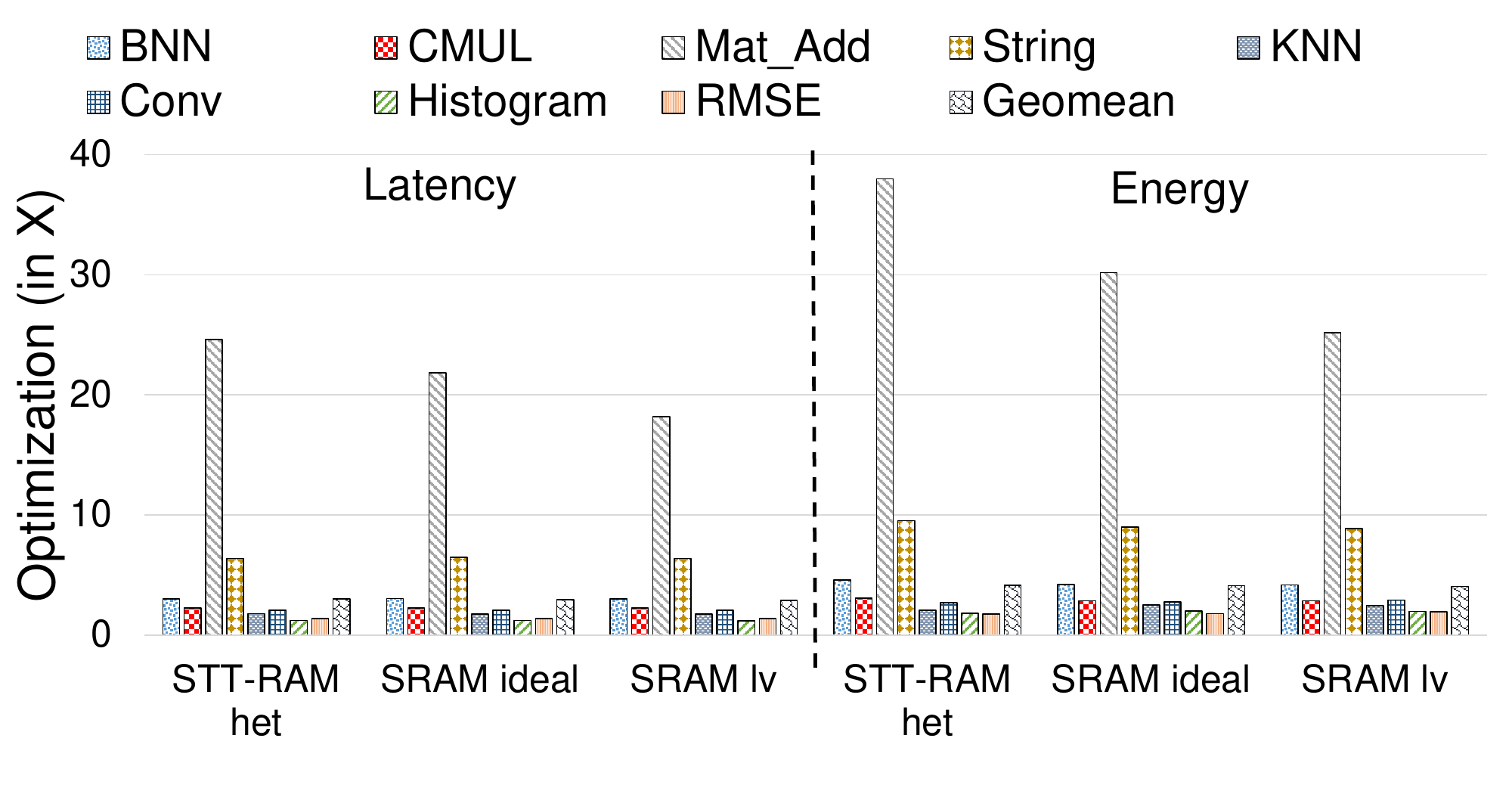}
  \vspace{-10pt}
  \caption{L2 cache}
  \label{fig:l2}
\end{subfigure}
\vspace{-15pt}
\caption{L1/L2 STT-RAM, ideal SRAM (SRAM$_{ideal}$) and low voltage word-line SRAM (SRAM$_{lv}$) compared to CPU.}
\label{fig:stt_sram}
\vspace{-15pt}
\end{figure*}

\subsubsection{L1 cache}
For the L1 cache, we use a homogeneous STT-RAM with a 75 $\mu$s retention time. 
Figure \ref{fig:l1} depicts the optimization achieved using different L1 PiC implementations compared to CPU-only computing for all the workloads. On average, STT-RAM, SRAM$_{ideal}$ and SRAM$_{lv}$ reduced the latency by 2.27x, 2.3x, and 2.22x, respectively, compared to the CPU. For all of our applications, SRAM$_{ideal}$ had the best execution time due to low write overheads compared to STT-RAM. Although the add operations in STT-RAM had fewer cycles than SRAM$_{ideal}$, as seen in Table \ref{tab:cycles}, it was not enough to offset the write overheads caused by STT-RAMs. Note, however, that SRAM$_{ideal}$ is a theoretical design and not realistic in practice due to data corruption issues in SRAM \cite{jeloka201628}. We observed that the highest latency optimizations were achieved for $Mat\_add$ on L1 caches by STT-RAM (6.59x), SRAM$_{ideal}$ (6.82x), and SRAM$_{lv}$ (5.79x).
 Compared to STT-RAM, SRAM$_{lv}$ slightly increased the latency by an average of 1.95\% and by up to 12\% for $Mat\_add$ due to longer access latencies of SRAM$_{lv}$. 

STT-RAM PiC performed much better with respect to energy. STT-RAM, SRAM$_{ideal}$, and SRAM$_{lv}$ reduced the energy by 2.84x, 2.72x, and 2.68x, respectively, compared to CPU. STT-RAM outperformed SRAM$_{ideal}$ by 4.56\%. Although SRAM$_{lv}$ had a lower dynamic power than SRAM$_{ideal}$, it had higher total energy because it ran slower. We observed the highest improvement for matrix addition ($Mat\_add$), which exhibited high data reuse---and more computations per data movement---during sum accumulation. The L1 cache reduced the latency of $Mat\_add$ by 6.58x, 6.819x, and 5.79x,  and reduced the energy by 10.19x, 7.76x, and 6.59x using STT-RAM, SRAM$_{ideal}$, and SRAM$_{lv}$, respectively. However, CPU-dependent workloads had the lowest optimizations due to high amounts of data movement (55\% - 75\% of total execution) and smaller workload portions running on PiC/PiM. The latency improved for the CPU-dependent workloads by an average of 1.43x, 1.44x, and 1.42x, and energy was reduced by 1.73x, 1.78x, and 1.84x using STT-RAM, SRAM $_{ideal}$, and SRAM$_{lv}$ respectively.

\subsubsection{L2 cache}
As mentioned earlier, the STT-RAM$_{het}$ is used in the L2 cache and features a 10ms retention time for CPU-based computing and 75$\mu$s retention time for PiC-based computing.
When comparing L1 and L2 caches, we observed higher optimization in L2 PiC than in L1 PiC due to more parallel units and lower data movement overhead from memory. As seen in Figure \ref{fig:l2}, STT-RAM$_{het}$, SRAM$_{ideal}$ and SRAM$_{lv}$ reduced the latency by 3x, 2.96x, and 2.889x, respectively, compared to the CPU. STT-RAMs achieved higher speedup than SRAM for ADD operations due to faster read operations, enabling longer slack to perform more bit additions per cycle. SRAM$_{ideal}$ achieved the fastest speedup for logical instructions, as seen in $BNN$, $CMUL$, and $String$ applications due to SRAM's faster write operations. For these applications, STT-RAM$_{het}$ only degraded performance by 0.68\%, 0.35\% and 1.15\%, respectively. However, STT-RAM$_{het}$ had better latency for ADD operation than SRAM due to better hit-latency slack. This allowed STT-RAM ADD operations to ripple-carry more bits and thus have better latency than SRAMs. This resulted in overall performance benefits of 1.52\% over SRAM$_{ideal}$. SRAM$_{lv}$ achieved a low speedup for logical and ADD instructions. Compared to SRAM$_{lv}$, STT-RAM$_{het}$ reduced the average latency by 4.21\% and up to 35.26\% for $Mat\_add$ application.


The energy savings from L2 PiC were similarly quite substantial, as seen in Figure \ref{fig:l2}. Energy savings compared to CPU-only computing for STT-RAM$_{het}$, SRAM$_{ideal}$ and SRAM$_{lv}$ were 4.16x, 4.12x, and 4.05x, respectively. The energy savings were highest for $Mat\_add$: 37.9x, 30.19x, and 25.18x with STT-RAM$_{het}$, SRAM$_{ideal}$, and SRAM$_{lv}$, respectively.
SRAM$_{ideal}$ outperformed STT-RAM$_{het}$ for CPU-dependent workloads by 8.4\$ on average because of the high write energy overheads incurred by the STT-RAM due to a high amount of data movements from the L2 cache to the CPU and vice-versa. 
However, in general, STT-RAM$_{het}$ resulted in better optimization than SRAM$_{ideal}$ by an average of 1.14\% and achieved the highest optimization for $Mat\_add$ by 25.85\%. 

STT-RAM$_{het}$ performed better than SRAM$_{low}$ by an average of 2.70\%, indicating that the static energy overhead of slower SRAM$_{low}$ is higher than the dynamic energy overhead of STT-RAMs. Moreover, using a reliable SRAM$_{low}$ for PiC computations will also degrade the standard cache operations for CPU-based computing due to their slower latency while simultaneously increasing the area compared to STT-RAMs. 

\begin{figure*}[ht]
\centering
\begin{subfigure}{.5\linewidth}
  \centering
  \includegraphics[width=.69\linewidth]{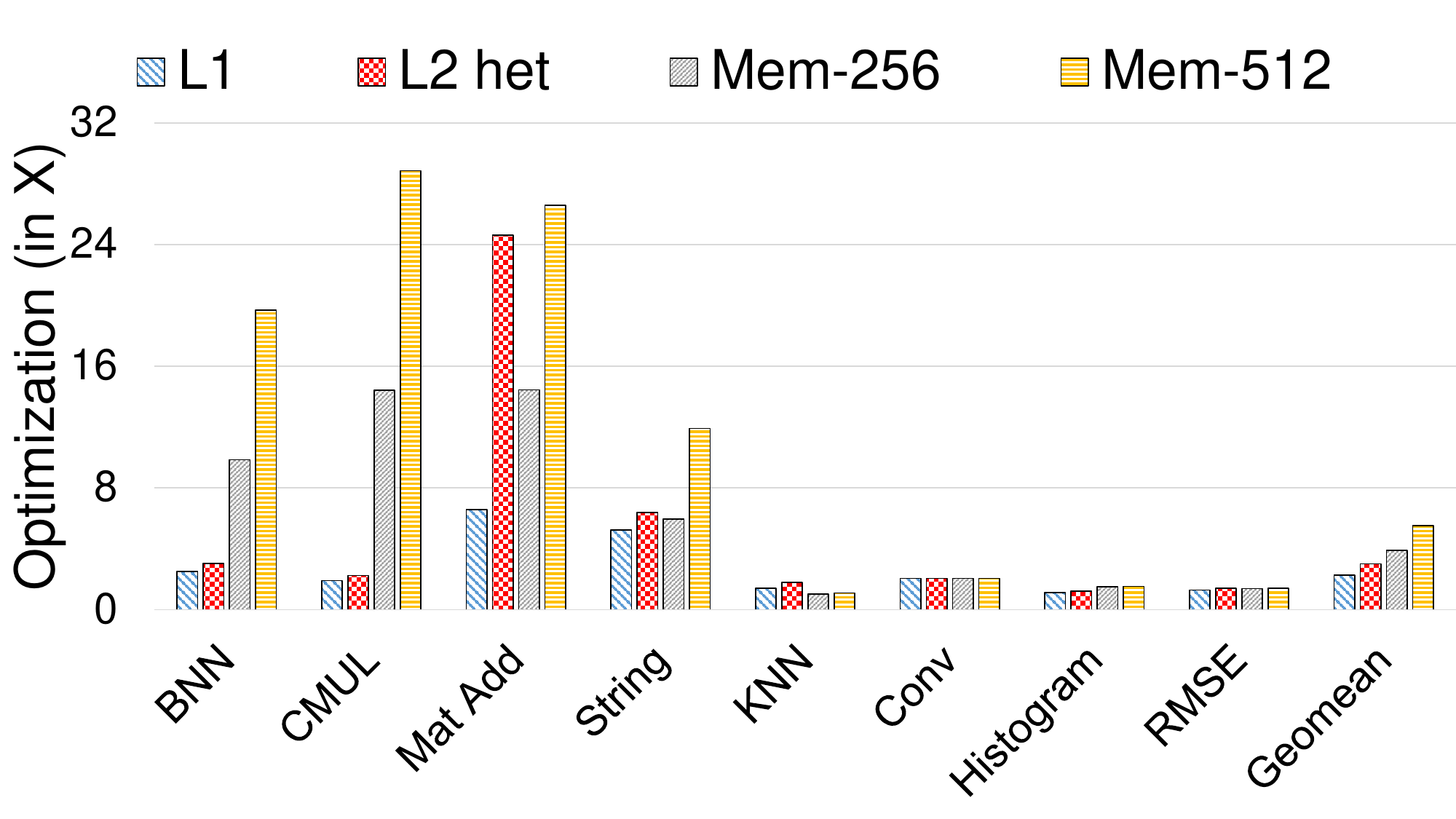}
  \vspace{-10pt}
  \caption{Latency}
  \label{fig:mem_lat}
\end{subfigure}%
\begin{subfigure}{.5\linewidth}
  \centering
  \includegraphics[width=.69\linewidth]{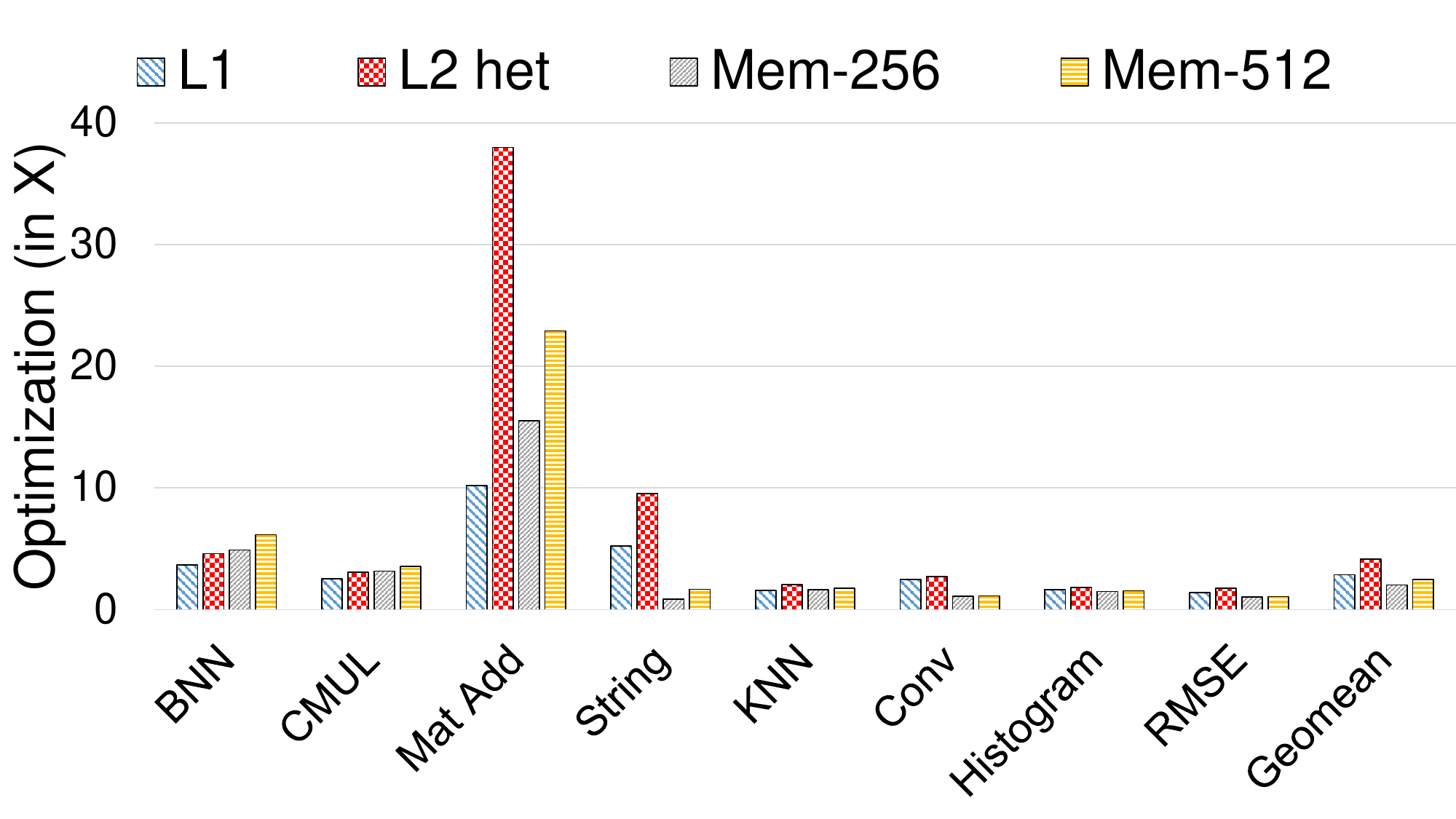}
  \vspace{-10pt}
  \caption{Energy}
  \label{fig:mem_ener}
\end{subfigure}
\caption{Relaxed retention STT-RAM PiC and non-volatile STT-RAM PiM compared to CPU. The memory that does 256 and 512 integer computations is called \textit{Mem-256} and \textit{Mem-512}.}
\label{fig:mem_Cache}
\vspace{-15pt}
\end{figure*}

\subsection{PiC vs. PiM}
We explore the best level of the hierarchy for  STT-RAM-based PiC/PiM. For these experiments, we compare PiC with 75$\mu$s retention time for L1 cache, heterogeneous STT-RAM configuration for L2 cache, referred to as \textit{L2$_{het}$}, and non-volatile PiM. The L1 cache supports 16 32-bit computations, L2 supports 64 32-bit computations, and we explored memory with the capability for 256 or 512 32-bit computations (called \textit{Mem-256} and \textit{Mem-512}, respectively). Similar to prior subsections, the results are normalized to CPU-based computing featuring STT-RAM caches at L1 and L2 levels.

Figure \ref{fig:mem_Cache} presents the latency and energy results for computing across the memory hierarchy. As seen in Figure \ref{fig:mem_lat}, L1 cache, L2$_{het}$, Mem-256, and Mem-512 improved the latency by 2.26x, 3x, 3.90x, and 5.53x, respectively, compared to CPU-only computing. A closer look at the workloads showed that different parts of the hierarchy were preferred for different types of workloads.
The average improvement for CPU-dependent workloads was 1.46x, 1.58x, 1.45x, and 1.49x for the L1 cache, L2$_{het}$, Mem-256, and Mem-512, respectively. L2 PiC achieved the highest optimization for CPU-dependent workloads due to the reduced data movement overhead from the processor and a large number of parallel units. However, for CPU-independent workloads, PiM had the fastest execution due to low data movement and a large number of parallel units. We observed a speedup of 3.58, 5.72x, 10.52x, and 20.59x for CPU-independent workloads, respectively. This shows that PiC works best for CPU-dependent workloads, whereas PiM works best for CPU-independent workloads despite having high access latency costs, as seen in Table \ref{tab:cycles}. 

Contrary to latency, the L2 PiC outperformed the PiM in energy savings (Figure \ref{fig:mem_ener}). On average, the L1 cache, L2$_{het}$, Mem-256, and Mem-512 reduced the energy by 2.86x, 4.16x, 2.2x, and 2.69x, respectively, compared to CPU. Thus, L2 caches are more energy-efficient than Mem-256 and Mem-512 due to reduced data movement and lower access latencies and write overheads. For CPU-dependent workloads L1 cache, L2$_{het}$, L2$_{10ms}$, Mem-256, and Mem-512 improved energy by 1.73x, 2.05x, 1.28x, and 1.34x, respectively, and by 4.72x, 8.45x, 8.07x, 3.77x, and 5.39x for CPU-independent workloads.

We also observed that even though PiM's and PiC's execution latencies could be hidden behind operation chaining, the number of computations that are performed remained the same. This puts PiM at a disadvantage due to very high write-energy overheads. Furthermore, high data reuse workloads also require additional writes within the same memory hierarchy level, which greatly increases the dynamic energy for the non-volatile memory and favors PiC by increasing the computations per data transferred. PiM only achieved better energy savings for two applications, $BNN$ and $CMUL$, both of which mainly perform logical operations and feature few write operations. Compared to L2$_{het}$, Mem-256 and Mem-512 optimized energy by 5.94\% and 25.08\% for $BNN$ and 2.90\% and 14.07\% for $CMUL$.

\subsection{Overhead}

The implemented PiC/PiM compute units had a critical path duration of 120ps and did not introduce any additional latency to cache access.
The energy consumed per bit for logical and ADD operations were 0.6pJ and 1.04pJ, respectively. In the case of STT-RAM L1, STT-RAM L2, and STT-RAM memory, the compute elements accounted for only 11.16\%, 3.54\%, and 0.55\% of the total energy consumption. For a 512x512 subarray of 6T SRAM, the compute unit occupied 3.7\% of the total area. With STT-RAMs being twice as dense as SRAM, our compute units required 6.37\% of the subarray area. In terms of cache area, the STT-RAM cache was 43.47\% and 79.86\% smaller than SRAM for L1 and homogeneous L2 cache, respectively. The heterogeneous L2 architecture required approximately 90\% more area than the homogeneous L2 cache design but still consumed 61.75\% less area than an SRAM L2 cache.

These area savings achieved by STT-RAMs provide more flexibility to incorporate more complex computational units in systems with limited resources. Moreover, PiC based on SRAM incurs additional performance overheads to mitigate data corruption issues \cite{jeloka201628}, which can also slow down CPU-based computing. In general, STT-RAMs offer a more robust and low-overhead solution for PiC implementations, characterized by lower area requirements and minimal overheads in preventing data corruption issues.


\section{Open research challenges}

Bit-line computing situates compute units within memories to increase parallelism and reduce data movement overhead. Despite advancements in this computing paradigm, several research challenges persist in bit-line computing, which also applies to hierarchical in-memory computing as described in this paper. Bit-line computing currently presents limitations due to its focus on domain-specific problems and the necessity for careful data alignment. To perform computations, data must be bit-aligned within the memory array. This specialization favors array-based workloads such as matrix operations commonly found in machine learning and image processing. Consequently, bitline computing excels in domain-specific architectures equipped with specialized compute units tailored for these aligned computations. Additionally, chiplet-based computing solutions can adopt bit-line computing accelerators as modularized PiC/PiM chiplets. These chiplets can be optimized for the specific bit-aligned workloads where this technique would be most impactful.

Another prominent challenge is the exploration of an instruction set architecture for PiC/PiM to enable more computations per instruction. For instance, the instruction table in Figure \ref{fig:het_arch} stores the addresses of 32-bit instructions and can be further optimized to store the address of an entire block within the subarray, facilitating specific bit-aligned operations. Thus, a single entry in the instruction table can point to two cache blocks to perform 16 32-bit computations (assuming a cache block size of 64B) on bit-aligned data.

Compiler capabilities need to be enhanced to identify the regions of code that are PiC/PiM friendly. This will allow users to run legacy code with PiC/PiM optimization without modifying the code. Such modifications can be inspired by the compiler handling of SIMD instructions, where the compiler first scans the code to determine its vectorizability and then computes a cost function to estimate the benefits of running the code on a SIMD architecture. If the conditions are met, the code is converted into SIMD executable code. Similarly, for PiC/PiM computing, compilers can check if the code offers parallelism and will not have significant data movements to diminish the advantages of PiC/PiM computing.
Furthermore, the management of the cache and memory controller capabilities in handling non-bit-aligned data or rearranging data within different sub-arrays needs exploration. One solution could be employing larger subarray sizes, but this would increase overall cache access latency to accommodate more data within the same sub-array. Alternatively, enhancing cache controller capabilities to manage data transfer within the cache and conceal data movement with the cache as much as possible would be beneficial.

Hardware exploration should include error correction techniques for STT-RAM-based PiC/PiM computing, tradeoff analysis of sub-array size, cache access latencies, number of compute units, and the overall speedup offered by PiC computing. Additionally, automating the PiC/PiM compute unit design flow to optimize complex compute unit designs with low area, latency, and power overheads on the cache and memories is crucial. However, larger compute units would result in higher routing overheads in the caches and memories, necessitating a cache/memory-aware compute unit design process.
Finally, increasing the number of compute units in the cache leads to a higher number of cache reads/writes, significantly increasing cache power consumption. Thus, studying the impact of thermal constraints arising from PiC/PiM computing to establish limitations on the number of active compute units is essential.

\section{Conclusion}

In this article, we conducted the first study of the STT-RAM cache as a candidate for processing in cache (PiC). We compared relaxed retention STT-RAM PiC to SRAM. Also, we compared STT-RAM PiC with non-volatile STT-RAM processing in memory (PiM) to analyze the tradeoffs of computing at different levels of the memory hierarchy. Additionally, we analyzed homogeneous and heterogeneous STT-RAM cache architectures featuring different retention times for CPU and PiC computing. For our analysis, we explored three types of workloads: CPU-dependent, CPU-independent with low data reuse, and CPU-independent with high data reuse. Our analysis revealed that STT-RAM offers an excellent opportunity for energy- and area-efficient PiC, providing latency benefits similar to those of SRAM. We also found that the characteristics of the executing workloads influence the choice between PiC and PiM. For instance, STT-RAM PiC outperforms PiM regarding latency optimization for CPU-dependent workloads with low instruction-level parallelism (ILP). This study demonstrates the significant promise of STT-RAM-based PiC and calls for further research to effectively implement it in emerging resource-constrained systems.

Future work involves exploring solutions to some of the open research challenges in the area of hierarchical in-memory computing. For instance, we plan to explore data-flow-like computer architectures using STT-RAM hierarchical computing, where different levels of the memory hierarchy have distinct sets of compute units. This approach allows the entire computer system to leverage a wider range of compute capabilities to enhance the system's overall performance and energy efficiency. Additionally, we plan to explore design space exploration strategies for these architectures. Using effective exploration strategies will enable the systematic evaluation of various architectural configurations to identify Pareto-optimal solutions that balance multiple objectives, such as performance, data movement overhead minimization, power consumption, and area efficiency. 

Furthermore, we plan to explore the automation of the PiC/PiM compute unit design flow to optimize the compute unit design for low area, latency, energy, and energy-delay product. Given that the architectures explored herein are domain-specific, such a design automation will lead to more efficient and effective designs, reducing the time and effort required for development. We would also like to explore instruction set architecture for PiC/PiM, focusing on streamlining the bit-line alignment of data, thereby easing the burden on the cache and memory controllers. Optimizing the data alignment process will make the proposed PiC/PiM systems more viable for a wide range of computing applications. 

\section*{Acknowledgment}
This work was partly supported by the National Science Foundation (NSF) under grant CNS-1844952. Any views expressed in this material are those of the authors and not necessarily of the NSF.


\bibliographystyle{IEEEtran}
\bibliography{IEEEabrv, refs}


\vskip 0pt plus -1fil
\begin{IEEEbiography}
[{\includegraphics[width=1in,height=1.25in,clip,keepaspectratio]{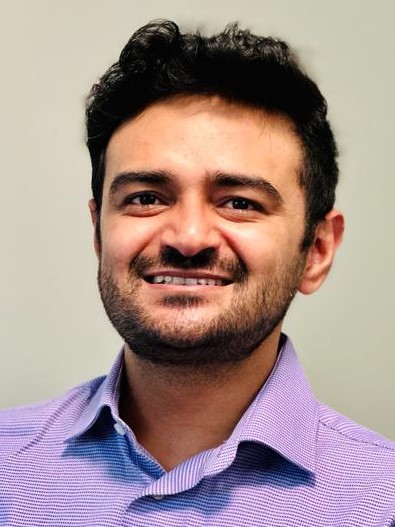}}]{Dhruv Gajaria} received his M.S. and Ph.D. in Electrical and Computer Engineering from the University of Arizona in 2019 and 2023, respectively, and his B.Eng in Electronics Engineering from the University of Mumbai, India, in 2017. He is a Post-Doctoral Research Associate with the High Performance Computing Group at Pacific Northwest National Lab, USA. His research interests include hardware-software co-design, computer architecture, simulation and performance analysis, and domain-specific architectures.

\end{IEEEbiography}
\vskip -20pt plus -1fil
\begin{IEEEbiography}
[{\includegraphics[width=1in,height=1.25in,clip,keepaspectratio]{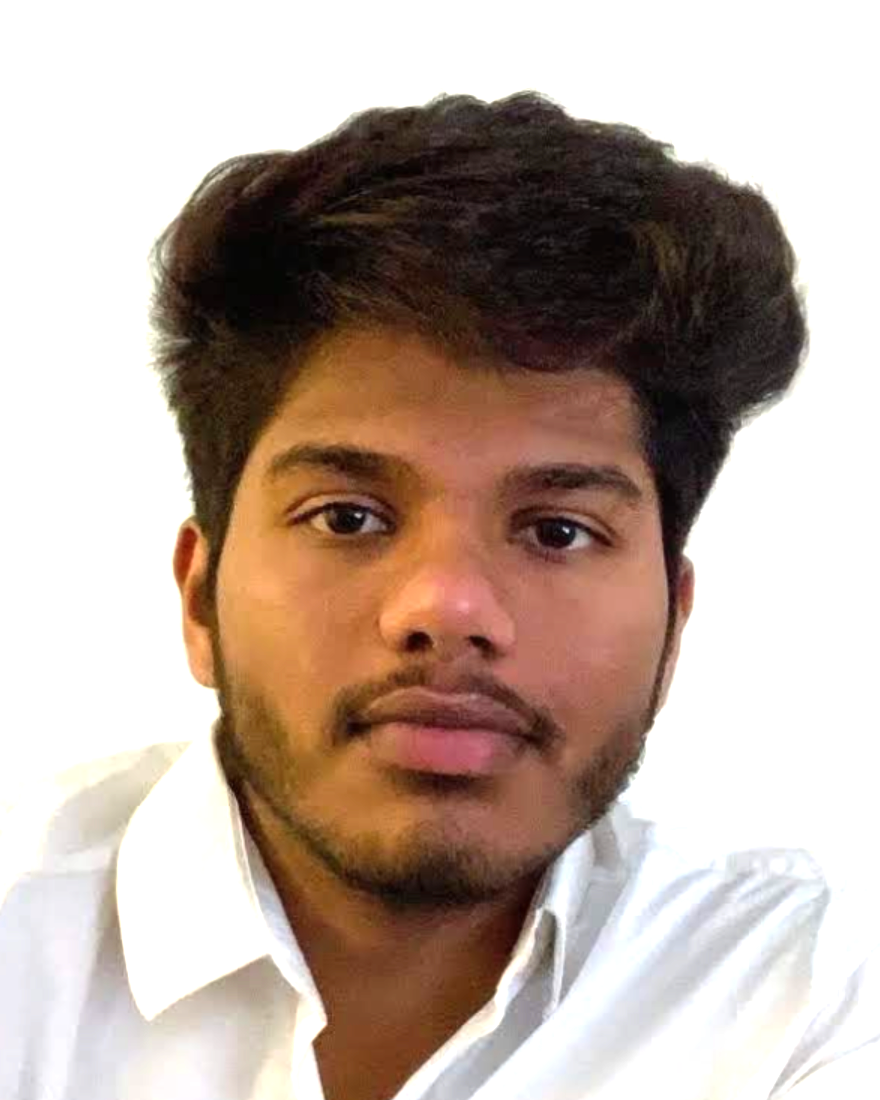}}]{Kevin Antony Gomez} received his B.S in Electrical and Computer Engineering from the University of Arizona in 2022. He is currently an M.S. student of Computer Science at the University of Massachusetts, Amherst. His research interests include in-cache/in-memory computing and ML tools for domain-specific architectures. 

\end{IEEEbiography}
\vskip -20pt plus -1fil
\begin{IEEEbiography}
[{\includegraphics[width=1in,height=1.25in,clip,keepaspectratio]{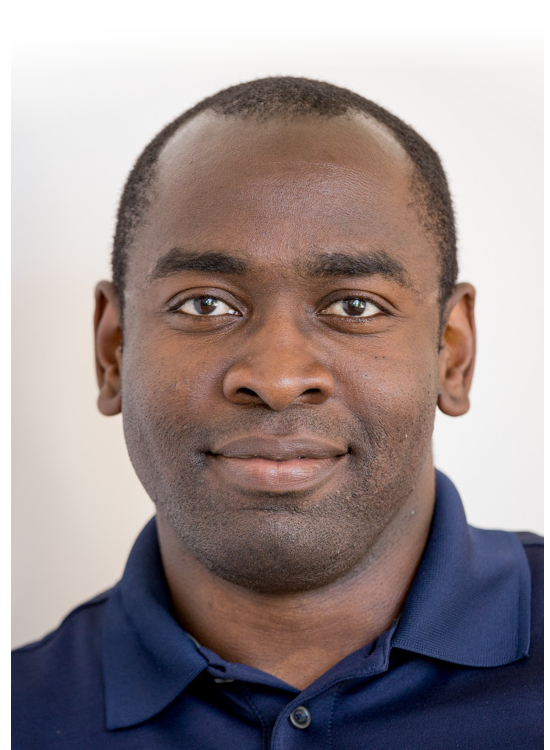}}]{Tosiron Adegbija} received his M.S and Ph.D. in Electrical and Computer Engineering from the University of Florida in 2011 and 2015, respectively and his B.Eng in Electrical Engineering from the University of Ilorin, Nigeria in 2005.
He is currently an Associate Professor of Electrical and Computer Engineering at the University of Arizona, USA. His research interests are in computer architecture, with an emphasis on brain-inspired computing, adaptable computing, low-power embedded systems design and optimization methodologies, and domain-specific architectures. He received the CAREER Award from the National Science Foundation in 2019. He currently serves as an Associate Editor for the IEEE Transactions on Computers and Embedded Systems Letters. He is a Senior Member of the IEEE.
\end{IEEEbiography}

\end{document}